\documentclass[longauth]{aaEC}

\usepackage{graphicx}
\usepackage{natbib}
\usepackage{scalerel}
\usepackage{comment}
\usepackage{booktabs}
\usepackage{longtable}
\usepackage{array}
\usepackage{soul}
\usepackage{float}

\usepackage[table]{xcolor}

\usepackage{txfonts}

\usepackage[pdfencoding=auto,psdextra]{hyperref}
\hypersetup{
    colorlinks=true,
    linkcolor=blue,
    filecolor=magenta,      
    urlcolor=blue,
    citecolor=blue
}

\makeatletter
\renewcommand*\aa@pageof{, page \thepage{} of \pageref*{LastPage}}
\makeatother

\usepackage[utf8]{inputenc}

\usepackage[switch, modulo]{lineno}
\linenumbers

\usepackage{euclid}

\usepackage{ulem}

\begin{document}

   \title{\Euclid\/: Asteroid rotation periods from the Euclid Ecliptic Survey\thanks{This paper is published on
     behalf of the Euclid Consortium.}}

\newcommand{\orcid}[1]{} 			   
\author{B.~Y.~Irureta-Goyena\orcid{0009-0004-5327-8767}\thanks{\email{belen.irureta@epfl.ch}}\inst{\ref{aff1}}
\and B.~Altieri\orcid{0000-0003-3936-0284}\inst{\ref{aff2}}
\and J.-P.~Kneib\orcid{0000-0002-4616-4989}\inst{\ref{aff1}}
\and M.~P\"ontinen\orcid{0000-0001-5442-2530}\inst{\ref{aff3}}
\and O.~R.~Hainaut\orcid{0000-0001-6952-9349}\inst{\ref{aff4}}
\and M.~R.~Alarcon\orcid{0000-0002-8134-2592}\inst{\ref{aff5}}
\and M.~Granvik\orcid{0000-0002-5624-1888}\inst{\ref{aff3},\ref{aff6}}
\and A.~A.~Nucita\inst{\ref{aff7},\ref{aff8},\ref{aff9}}
\and B.~Carry\orcid{0000-0001-5242-3089}\inst{\ref{aff10}}
\and M.~Devogele\orcid{0000-0002-6509-6360}\inst{\ref{aff11}}
\and M.~Mahlke\orcid{0000-0003-2831-0513}\inst{\ref{aff12}}
\and R.~Vavrek\inst{\ref{aff2}}
\and T.~M\"{u}ller\orcid{0000-0002-0717-0462}\inst{\ref{aff13}}
\and E.~Vilenius\orcid{0000-0002-6184-7681}\inst{\ref{aff14}}
\and C.~Snodgrass\orcid{0000-0001-9328-2905}\inst{\ref{aff15}}
\and R.~Kohley\inst{\ref{aff2}}
\and C.~Lemon\inst{\ref{aff16}}
\and P.~G\'omez-Alvarez\orcid{0000-0002-8594-5358}\inst{\ref{aff17},\ref{aff2}}
\and G.~Verdoes~Kleijn\orcid{0000-0001-5803-2580}\inst{\ref{aff18}}
\and J.~Licandro\orcid{0000-0002-9214-337X}\inst{\ref{aff5},\ref{aff19}}
\and S.~Kruk\orcid{0000-0001-8010-8879}\inst{\ref{aff2}}
\and L.~Conversi\orcid{0000-0002-6710-8476}\inst{\ref{aff11},\ref{aff2}}
\and A.~Franco\orcid{0000-0002-4761-366X}\inst{\ref{aff8},\ref{aff7},\ref{aff9}}
\and G.~Buenadicha\inst{\ref{aff2}}
\and P.~Mas-Buitrago\orcid{0000-0001-8055-7949}\inst{\ref{aff20}}
\and K.~Kuijken\orcid{0000-0002-3827-0175}\inst{\ref{aff21}}
\and S.~Andreon\orcid{0000-0002-2041-8784}\inst{\ref{aff22}}
\and C.~Baccigalupi\orcid{0000-0002-8211-1630}\inst{\ref{aff23},\ref{aff24},\ref{aff25},\ref{aff26}}
\and M.~Baldi\orcid{0000-0003-4145-1943}\inst{\ref{aff27},\ref{aff28},\ref{aff29}}
\and A.~Balestra\orcid{0000-0002-6967-261X}\inst{\ref{aff30}}
\and P.~Battaglia\orcid{0000-0002-7337-5909}\inst{\ref{aff28}}
\and A.~Biviano\orcid{0000-0002-0857-0732}\inst{\ref{aff24},\ref{aff23}}
\and E.~Branchini\orcid{0000-0002-0808-6908}\inst{\ref{aff31},\ref{aff32},\ref{aff22}}
\and M.~Brescia\orcid{0000-0001-9506-5680}\inst{\ref{aff33},\ref{aff34}}
\and S.~Camera\orcid{0000-0003-3399-3574}\inst{\ref{aff35},\ref{aff36},\ref{aff37}}
\and V.~Capobianco\orcid{0000-0002-3309-7692}\inst{\ref{aff37}}
\and C.~Carbone\orcid{0000-0003-0125-3563}\inst{\ref{aff38}}
\and J.~Carretero\orcid{0000-0002-3130-0204}\inst{\ref{aff39},\ref{aff40}}
\and R.~Casas\orcid{0000-0002-8165-5601}\inst{\ref{aff41},\ref{aff42}}
\and M.~Castellano\orcid{0000-0001-9875-8263}\inst{\ref{aff43}}
\and G.~Castignani\orcid{0000-0001-6831-0687}\inst{\ref{aff28}}
\and S.~Cavuoti\orcid{0000-0002-3787-4196}\inst{\ref{aff34},\ref{aff44}}
\and K.~C.~Chambers\orcid{0000-0001-6965-7789}\inst{\ref{aff45}}
\and A.~Cimatti\inst{\ref{aff46}}
\and C.~Colodro-Conde\inst{\ref{aff5}}
\and G.~Congedo\orcid{0000-0003-2508-0046}\inst{\ref{aff15}}
\and C.~J.~Conselice\orcid{0000-0003-1949-7638}\inst{\ref{aff47}}
\and Y.~Copin\orcid{0000-0002-5317-7518}\inst{\ref{aff48}}
\and F.~Courbin\orcid{0000-0003-0758-6510}\inst{\ref{aff49},\ref{aff50},\ref{aff51}}
\and H.~M.~Courtois\orcid{0000-0003-0509-1776}\inst{\ref{aff52}}
\and M.~Cropper\orcid{0000-0003-4571-9468}\inst{\ref{aff14}}
\and H.~Degaudenzi\orcid{0000-0002-5887-6799}\inst{\ref{aff53}}
\and G.~De~Lucia\orcid{0000-0002-6220-9104}\inst{\ref{aff24}}
\and C.~Dolding\orcid{0009-0003-7199-6108}\inst{\ref{aff14}}
\and H.~Dole\orcid{0000-0002-9767-3839}\inst{\ref{aff54}}
\and F.~Dubath\orcid{0000-0002-6533-2810}\inst{\ref{aff53}}
\and X.~Dupac\inst{\ref{aff2}}
\and M.~Farina\orcid{0000-0002-3089-7846}\inst{\ref{aff55}}
\and R.~Farinelli\inst{\ref{aff28}}
\and S.~Ferriol\inst{\ref{aff48}}
\and M.~Frailis\orcid{0000-0002-7400-2135}\inst{\ref{aff24}}
\and M.~Fumana\orcid{0000-0001-6787-5950}\inst{\ref{aff38}}
\and S.~Galeotta\orcid{0000-0002-3748-5115}\inst{\ref{aff24}}
\and K.~George\orcid{0000-0002-1734-8455}\inst{\ref{aff56}}
\and B.~Gillis\orcid{0000-0002-4478-1270}\inst{\ref{aff15}}
\and C.~Giocoli\orcid{0000-0002-9590-7961}\inst{\ref{aff28},\ref{aff29}}
\and J.~Gracia-Carpio\orcid{0000-0003-4689-3134}\inst{\ref{aff13}}
\and A.~Grazian\orcid{0000-0002-5688-0663}\inst{\ref{aff30}}
\and F.~Grupp\inst{\ref{aff13},\ref{aff57}}
\and S.~V.~H.~Haugan\orcid{0000-0001-9648-7260}\inst{\ref{aff58}}
\and H.~Hoekstra\orcid{0000-0002-0641-3231}\inst{\ref{aff21}}
\and W.~Holmes\inst{\ref{aff59}}
\and I.~M.~Hook\orcid{0000-0002-2960-978X}\inst{\ref{aff60}}
\and F.~Hormuth\inst{\ref{aff61}}
\and A.~Hornstrup\orcid{0000-0002-3363-0936}\inst{\ref{aff62},\ref{aff63}}
\and K.~Jahnke\orcid{0000-0003-3804-2137}\inst{\ref{aff64}}
\and M.~Jhabvala\inst{\ref{aff65}}
\and A.~Kiessling\orcid{0000-0002-2590-1273}\inst{\ref{aff59}}
\and B.~Kubik\orcid{0009-0006-5823-4880}\inst{\ref{aff48}}
\and M.~K\"ummel\orcid{0000-0003-2791-2117}\inst{\ref{aff57}}
\and M.~Kunz\orcid{0000-0002-3052-7394}\inst{\ref{aff66}}
\and H.~Kurki-Suonio\orcid{0000-0002-4618-3063}\inst{\ref{aff3},\ref{aff67}}
\and A.~M.~C.~Le~Brun\orcid{0000-0002-0936-4594}\inst{\ref{aff68}}
\and S.~Ligori\orcid{0000-0003-4172-4606}\inst{\ref{aff37}}
\and P.~B.~Lilje\orcid{0000-0003-4324-7794}\inst{\ref{aff58}}
\and V.~Lindholm\orcid{0000-0003-2317-5471}\inst{\ref{aff3},\ref{aff67}}
\and I.~Lloro\orcid{0000-0001-5966-1434}\inst{\ref{aff69}}
\and G.~Mainetti\orcid{0000-0003-2384-2377}\inst{\ref{aff70}}
\and O.~Mansutti\orcid{0000-0001-5758-4658}\inst{\ref{aff24}}
\and O.~Marggraf\orcid{0000-0001-7242-3852}\inst{\ref{aff71}}
\and M.~Martinelli\orcid{0000-0002-6943-7732}\inst{\ref{aff43},\ref{aff72}}
\and N.~Martinet\orcid{0000-0003-2786-7790}\inst{\ref{aff73}}
\and F.~Marulli\orcid{0000-0002-8850-0303}\inst{\ref{aff74},\ref{aff28},\ref{aff29}}
\and R.~J.~Massey\orcid{0000-0002-6085-3780}\inst{\ref{aff75}}
\and E.~Medinaceli\orcid{0000-0002-4040-7783}\inst{\ref{aff28}}
\and S.~Mei\orcid{0000-0002-2849-559X}\inst{\ref{aff76},\ref{aff77}}
\and E.~Merlin\orcid{0000-0001-6870-8900}\inst{\ref{aff43}}
\and G.~Meylan\inst{\ref{aff1}}
\and A.~Mora\orcid{0000-0002-1922-8529}\inst{\ref{aff78}}
\and L.~Moscardini\orcid{0000-0002-3473-6716}\inst{\ref{aff74},\ref{aff28},\ref{aff29}}
\and R.~Nakajima\orcid{0009-0009-1213-7040}\inst{\ref{aff71}}
\and C.~Neissner\orcid{0000-0001-8524-4968}\inst{\ref{aff79},\ref{aff40}}
\and S.-M.~Niemi\orcid{0009-0005-0247-0086}\inst{\ref{aff80}}
\and C.~Padilla\orcid{0000-0001-7951-0166}\inst{\ref{aff79}}
\and S.~Paltani\orcid{0000-0002-8108-9179}\inst{\ref{aff53}}
\and F.~Pasian\orcid{0000-0002-4869-3227}\inst{\ref{aff24}}
\and K.~Pedersen\inst{\ref{aff81}}
\and W.~J.~Percival\orcid{0000-0002-0644-5727}\inst{\ref{aff82},\ref{aff83},\ref{aff84}}
\and V.~Pettorino\orcid{0000-0002-4203-9320}\inst{\ref{aff80}}
\and G.~Polenta\orcid{0000-0003-4067-9196}\inst{\ref{aff85}}
\and L.~A.~Popa\inst{\ref{aff86}}
\and F.~Raison\orcid{0000-0002-7819-6918}\inst{\ref{aff13}}
\and R.~Rebolo\orcid{0000-0003-3767-7085}\inst{\ref{aff5},\ref{aff87},\ref{aff19}}
\and A.~Renzi\orcid{0000-0001-9856-1970}\inst{\ref{aff88},\ref{aff89},\ref{aff28}}
\and J.~Rhodes\orcid{0000-0002-4485-8549}\inst{\ref{aff59}}
\and G.~Riccio\inst{\ref{aff34}}
\and E.~Romelli\orcid{0000-0003-3069-9222}\inst{\ref{aff24}}
\and M.~Roncarelli\orcid{0000-0001-9587-7822}\inst{\ref{aff28}}
\and R.~Saglia\orcid{0000-0003-0378-7032}\inst{\ref{aff57},\ref{aff13}}
\and Z.~Sakr\orcid{0000-0002-4823-3757}\inst{\ref{aff90},\ref{aff91},\ref{aff92}}
\and D.~Sapone\orcid{0000-0001-7089-4503}\inst{\ref{aff93}}
\and M.~Schirmer\orcid{0000-0003-2568-9994}\inst{\ref{aff64}}
\and P.~Schneider\orcid{0000-0001-8561-2679}\inst{\ref{aff71}}
\and A.~Secroun\orcid{0000-0003-0505-3710}\inst{\ref{aff94}}
\and E.~Sihvola\orcid{0000-0003-1804-7715}\inst{\ref{aff95}}
\and P.~Simon\inst{\ref{aff71}}
\and C.~Sirignano\orcid{0000-0002-0995-7146}\inst{\ref{aff88},\ref{aff89}}
\and G.~Sirri\orcid{0000-0003-2626-2853}\inst{\ref{aff29}}
\and L.~Stanco\orcid{0000-0002-9706-5104}\inst{\ref{aff89}}
\and P.~Tallada-Cresp\'{i}\orcid{0000-0002-1336-8328}\inst{\ref{aff39},\ref{aff40}}
\and I.~Tereno\orcid{0000-0002-4537-6218}\inst{\ref{aff96},\ref{aff97}}
\and S.~Toft\orcid{0000-0003-3631-7176}\inst{\ref{aff98},\ref{aff99}}
\and R.~Toledo-Moreo\orcid{0000-0002-2997-4859}\inst{\ref{aff100}}
\and F.~Torradeflot\orcid{0000-0003-1160-1517}\inst{\ref{aff40},\ref{aff39}}
\and I.~Tutusaus\orcid{0000-0002-3199-0399}\inst{\ref{aff42},\ref{aff41},\ref{aff91}}
\and J.~Valiviita\orcid{0000-0001-6225-3693}\inst{\ref{aff3},\ref{aff67}}
\and T.~Vassallo\orcid{0000-0001-6512-6358}\inst{\ref{aff24},\ref{aff56}}
\and Y.~Wang\orcid{0000-0002-4749-2984}\inst{\ref{aff101}}
\and J.~Weller\orcid{0000-0002-8282-2010}\inst{\ref{aff57},\ref{aff13}}
\and F.~M.~Zerbi\orcid{0000-0002-9996-973X}\inst{\ref{aff22}}
\and J.~Garc\'ia-Bellido\orcid{0000-0002-9370-8360}\inst{\ref{aff90}}
\and J.~Mart\'{i}n-Fleitas\orcid{0000-0002-8594-569X}\inst{\ref{aff102}}
\and V.~Scottez\orcid{0009-0008-3864-940X}\inst{\ref{aff103},\ref{aff104}}
\and G.~Helou\orcid{0000-0003-3367-3415}\inst{\ref{aff105}}
\and D.~Scott\orcid{0000-0002-6878-9840}\inst{\ref{aff106}}}
										   
\institute{Institute of Physics, Laboratory of Astrophysics, Ecole Polytechnique F\'ed\'erale de Lausanne (EPFL), Observatoire de Sauverny, 1290 Versoix, Switzerland\label{aff1}
\and
ESAC/ESA, Camino Bajo del Castillo, s/n., Urb. Villafranca del Castillo, 28692 Villanueva de la Ca\~nada, Madrid, Spain\label{aff2}
\and
Department of Physics, P.O. Box 64, University of Helsinki, 00014 Helsinki, Finland\label{aff3}
\and
European Southern Observatory, Karl-Schwarzschild-Str.~2, 85748 Garching, Germany\label{aff4}
\and
Instituto de Astrof\'{\i}sica de Canarias, E-38205 La Laguna, Tenerife, Spain\label{aff5}
\and
Asteroid Engineering Laboratory, Lule\aa{} University of Technology, Box 848, 98128 Kiruna, Sweden\label{aff6}
\and
Department of Mathematics and Physics E. De Giorgi, University of Salento, Via per Arnesano, CP-I93, 73100, Lecce, Italy\label{aff7}
\and
INFN, Sezione di Lecce, Via per Arnesano, CP-193, 73100, Lecce, Italy\label{aff8}
\and
INAF-Sezione di Lecce, c/o Dipartimento Matematica e Fisica, Via per Arnesano, 73100, Lecce, Italy\label{aff9}
\and
Universit\'e C\^{o}te d'Azur, Observatoire de la C\^{o}te d'Azur, CNRS, Laboratoire Lagrange, Bd de l'Observatoire, CS 34229, 06304 Nice cedex 4, France\label{aff10}
\and
European Space Agency/ESRIN, Largo Galileo Galilei 1, 00044 Frascati, Roma, Italy\label{aff11}
\and
Universit\'e de Franche-Comt\'e, Institut UTINAM, CNRS UMR6213, OSU THETA Franche-Comt\'e-Bourgogne, Observatoire de Besan\c con, BP 1615, 25010 Besan\c con Cedex, France\label{aff12}
\and
Max Planck Institute for Extraterrestrial Physics, Giessenbachstr. 1, 85748 Garching, Germany\label{aff13}
\and
Mullard Space Science Laboratory, University College London, Holmbury St Mary, Dorking, Surrey RH5 6NT, UK\label{aff14}
\and
Institute for Astronomy, University of Edinburgh, Royal Observatory, Blackford Hill, Edinburgh EH9 3HJ, UK\label{aff15}
\and
Oskar Klein Centre for Cosmoparticle Physics, Department of Physics, Stockholm University, Stockholm, SE-106 91, Sweden\label{aff16}
\and
FRACTAL S.L.N.E., calle Tulip\'an 2, Portal 13 1A, 28231, Las Rozas de Madrid, Spain\label{aff17}
\and
Kapteyn Astronomical Institute, University of Groningen, PO Box 800, 9700 AV Groningen, The Netherlands\label{aff18}
\and
Universidad de La Laguna, Dpto. Astrof\'\i sica, E-38206 La Laguna, Tenerife, Spain\label{aff19}
\and
Centro de Astrobiolog\'ia (CAB), CSIC-INTA, ESAC Campus, Camino Bajo del Castillo s/n, 28692 Villanueva de la Ca\~nada, Madrid, Spain\label{aff20}
\and
Leiden Observatory, Leiden University, Einsteinweg 55, 2333 CC Leiden, The Netherlands\label{aff21}
\and
INAF-Osservatorio Astronomico di Brera, Via Brera 28, 20122 Milano, Italy\label{aff22}
\and
IFPU, Institute for Fundamental Physics of the Universe, via Beirut 2, 34151 Trieste, Italy\label{aff23}
\and
INAF-Osservatorio Astronomico di Trieste, Via G. B. Tiepolo 11, 34143 Trieste, Italy\label{aff24}
\and
INFN, Sezione di Trieste, Via Valerio 2, 34127 Trieste TS, Italy\label{aff25}
\and
SISSA, International School for Advanced Studies, Via Bonomea 265, 34136 Trieste TS, Italy\label{aff26}
\and
Dipartimento di Fisica e Astronomia, Universit\`a di Bologna, Via Gobetti 93/2, 40129 Bologna, Italy\label{aff27}
\and
INAF-Osservatorio di Astrofisica e Scienza dello Spazio di Bologna, Via Piero Gobetti 93/3, 40129 Bologna, Italy\label{aff28}
\and
INFN-Sezione di Bologna, Viale Berti Pichat 6/2, 40127 Bologna, Italy\label{aff29}
\and
INAF-Osservatorio Astronomico di Padova, Via dell'Osservatorio 5, 35122 Padova, Italy\label{aff30}
\and
Dipartimento di Fisica, Universit\`a di Genova, Via Dodecaneso 33, 16146, Genova, Italy\label{aff31}
\and
INFN-Sezione di Genova, Via Dodecaneso 33, 16146, Genova, Italy\label{aff32}
\and
Department of Physics "E. Pancini", University Federico II, Via Cinthia 6, 80126, Napoli, Italy\label{aff33}
\and
INAF-Osservatorio Astronomico di Capodimonte, Via Moiariello 16, 80131 Napoli, Italy\label{aff34}
\and
Dipartimento di Fisica, Universit\`a degli Studi di Torino, Via P. Giuria 1, 10125 Torino, Italy\label{aff35}
\and
INFN-Sezione di Torino, Via P. Giuria 1, 10125 Torino, Italy\label{aff36}
\and
INAF-Osservatorio Astrofisico di Torino, Via Osservatorio 20, 10025 Pino Torinese (TO), Italy\label{aff37}
\and
INAF-IASF Milano, Via Alfonso Corti 12, 20133 Milano, Italy\label{aff38}
\and
Centro de Investigaciones Energ\'eticas, Medioambientales y Tecnol\'ogicas (CIEMAT), Avenida Complutense 40, 28040 Madrid, Spain\label{aff39}
\and
Port d'Informaci\'{o} Cient\'{i}fica, Campus UAB, C. Albareda s/n, 08193 Bellaterra (Barcelona), Spain\label{aff40}
\and
Institut d'Estudis Espacials de Catalunya (IEEC),  Edifici RDIT, Campus UPC, 08860 Castelldefels, Barcelona, Spain\label{aff41}
\and
Institute of Space Sciences (ICE, CSIC), Campus UAB, Carrer de Can Magrans, s/n, 08193 Barcelona, Spain\label{aff42}
\and
INAF-Osservatorio Astronomico di Roma, Via Frascati 33, 00078 Monteporzio Catone, Italy\label{aff43}
\and
INFN section of Naples, Via Cinthia 6, 80126, Napoli, Italy\label{aff44}
\and
Institute for Astronomy, University of Hawaii, 2680 Woodlawn Drive, Honolulu, HI 96822, USA\label{aff45}
\and
Dipartimento di Fisica e Astronomia "Augusto Righi" - Alma Mater Studiorum Universit\`a di Bologna, Viale Berti Pichat 6/2, 40127 Bologna, Italy\label{aff46}
\and
Jodrell Bank Centre for Astrophysics, Department of Physics and Astronomy, University of Manchester, Oxford Road, Manchester M13 9PL, UK\label{aff47}
\and
Universit\'e Claude Bernard Lyon 1, CNRS/IN2P3, IP2I Lyon, UMR 5822, Villeurbanne, F-69100, France\label{aff48}
\and
Institut de Ci\`{e}ncies del Cosmos (ICCUB), Universitat de Barcelona (IEEC-UB), Mart\'{i} i Franqu\`{e}s 1, 08028 Barcelona, Spain\label{aff49}
\and
Instituci\'o Catalana de Recerca i Estudis Avan\c{c}ats (ICREA), Passeig de Llu\'{\i}s Companys 23, 08010 Barcelona, Spain\label{aff50}
\and
Institut de Ciencies de l'Espai (IEEC-CSIC), Campus UAB, Carrer de Can Magrans, s/n Cerdanyola del Vall\'es, 08193 Barcelona, Spain\label{aff51}
\and
UCB Lyon 1, CNRS/IN2P3, IUF, IP2I Lyon, 4 rue Enrico Fermi, 69622 Villeurbanne, France\label{aff52}
\and
Department of Astronomy, University of Geneva, ch. d'Ecogia 16, 1290 Versoix, Switzerland\label{aff53}
\and
Universit\'e Paris-Saclay, CNRS, Institut d'astrophysique spatiale, 91405, Orsay, France\label{aff54}
\and
INAF-Istituto di Astrofisica e Planetologia Spaziali, via del Fosso del Cavaliere, 100, 00100 Roma, Italy\label{aff55}
\and
University Observatory, LMU Faculty of Physics, Scheinerstr.~1, 81679 Munich, Germany\label{aff56}
\and
Universit\"ats-Sternwarte M\"unchen, Fakult\"at f\"ur Physik, Ludwig-Maximilians-Universit\"at M\"unchen, Scheinerstr.~1, 81679 M\"unchen, Germany\label{aff57}
\and
Institute of Theoretical Astrophysics, University of Oslo, P.O. Box 1029 Blindern, 0315 Oslo, Norway\label{aff58}
\and
Jet Propulsion Laboratory, California Institute of Technology, 4800 Oak Grove Drive, Pasadena, CA, 91109, USA\label{aff59}
\and
Department of Physics, Lancaster University, Lancaster, LA1 4YB, UK\label{aff60}
\and
Felix Hormuth Engineering, Goethestr. 17, 69181 Leimen, Germany\label{aff61}
\and
Technical University of Denmark, Elektrovej 327, 2800 Kgs. Lyngby, Denmark\label{aff62}
\and
Cosmic Dawn Center (DAWN), Denmark\label{aff63}
\and
Max-Planck-Institut f\"ur Astronomie, K\"onigstuhl 17, 69117 Heidelberg, Germany\label{aff64}
\and
NASA Goddard Space Flight Center, Greenbelt, MD 20771, USA\label{aff65}
\and
Universit\'e de Gen\`eve, D\'epartement de Physique Th\'eorique and Centre for Astroparticle Physics, 24 quai Ernest-Ansermet, CH-1211 Gen\`eve 4, Switzerland\label{aff66}
\and
Helsinki Institute of Physics, Gustaf H{\"a}llstr{\"o}min katu 2, University of Helsinki, 00014 Helsinki, Finland\label{aff67}
\and
Laboratoire d'etude de l'Univers et des phenomenes eXtremes, Observatoire de Paris, Universit\'e PSL, Sorbonne Universit\'e, CNRS, 92190 Meudon, France\label{aff68}
\and
SKAO, Jodrell Bank, Lower Withington, Macclesfield SK11 9FT, UK\label{aff69}
\and
Centre de Calcul de l'IN2P3/CNRS, 21 avenue Pierre de Coubertin 69627 Villeurbanne Cedex, France\label{aff70}
\and
Universit\"at Bonn, Argelander-Institut f\"ur Astronomie, Auf dem H\"ugel 71, 53121 Bonn, Germany\label{aff71}
\and
INFN-Sezione di Roma, Piazzale Aldo Moro, 2 - c/o Dipartimento di Fisica, Edificio G. Marconi, 00185 Roma, Italy\label{aff72}
\and
Aix-Marseille Universit\'e, CNRS, CNES, LAM, Marseille, France\label{aff73}
\and
Dipartimento di Fisica e Astronomia "Augusto Righi" - Alma Mater Studiorum Universit\`a di Bologna, via Piero Gobetti 93/2, 40129 Bologna, Italy\label{aff74}
\and
Department of Physics, Institute for Computational Cosmology, Durham University, South Road, Durham, DH1 3LE, UK\label{aff75}
\and
Universit\'e Paris Cit\'e, CNRS, Astroparticule et Cosmologie, 75013 Paris, France\label{aff76}
\and
CNRS-UCB International Research Laboratory, Centre Pierre Bin\'etruy, IRL2007, CPB-IN2P3, Berkeley, USA\label{aff77}
\and
Telespazio UK S.L. for European Space Agency (ESA), Camino bajo del Castillo, s/n, Urbanizacion Villafranca del Castillo, Villanueva de la Ca\~nada, 28692 Madrid, Spain\label{aff78}
\and
Institut de F\'{i}sica d'Altes Energies (IFAE), The Barcelona Institute of Science and Technology, Campus UAB, 08193 Bellaterra (Barcelona), Spain\label{aff79}
\and
European Space Agency/ESTEC, Keplerlaan 1, 2201 AZ Noordwijk, The Netherlands\label{aff80}
\and
DARK, Niels Bohr Institute, University of Copenhagen, Jagtvej 155, 2200 Copenhagen, Denmark\label{aff81}
\and
Waterloo Centre for Astrophysics, University of Waterloo, Waterloo, Ontario N2L 3G1, Canada\label{aff82}
\and
Department of Physics and Astronomy, University of Waterloo, Waterloo, Ontario N2L 3G1, Canada\label{aff83}
\and
Perimeter Institute for Theoretical Physics, Waterloo, Ontario N2L 2Y5, Canada\label{aff84}
\and
Space Science Data Center, Italian Space Agency, via del Politecnico snc, 00133 Roma, Italy\label{aff85}
\and
Institute of Space Science, Str. Atomistilor, nr. 409 M\u{a}gurele, Ilfov, 077125, Romania\label{aff86}
\and
Consejo Superior de Investigaciones Cientificas, Calle Serrano 117, 28006 Madrid, Spain\label{aff87}
\and
Dipartimento di Fisica e Astronomia "G. Galilei", Universit\`a di Padova, Via Marzolo 8, 35131 Padova, Italy\label{aff88}
\and
INFN-Padova, Via Marzolo 8, 35131 Padova, Italy\label{aff89}
\and
Instituto de F\'isica Te\'orica UAM-CSIC, Campus de Cantoblanco, 28049 Madrid, Spain\label{aff90}
\and
Institut de Recherche en Astrophysique et Plan\'etologie (IRAP), Universit\'e de Toulouse, CNRS, UPS, CNES, 14 Av. Edouard Belin, 31400 Toulouse, France\label{aff91}
\and
Universit\'e St Joseph; Faculty of Sciences, Beirut, Lebanon\label{aff92}
\and
Departamento de F\'isica, FCFM, Universidad de Chile, Blanco Encalada 2008, Santiago, Chile\label{aff93}
\and
Aix-Marseille Universit\'e, CNRS/IN2P3, CPPM, Marseille, France\label{aff94}
\and
Department of Physics and Helsinki Institute of Physics, Gustaf H\"allstr\"omin katu 2, University of Helsinki, 00014 Helsinki, Finland\label{aff95}
\and
Departamento de F\'isica, Faculdade de Ci\^encias, Universidade de Lisboa, Edif\'icio C8, Campo Grande, PT1749-016 Lisboa, Portugal\label{aff96}
\and
Instituto de Astrof\'isica e Ci\^encias do Espa\c{c}o, Faculdade de Ci\^encias, Universidade de Lisboa, Tapada da Ajuda, 1349-018 Lisboa, Portugal\label{aff97}
\and
Cosmic Dawn Center (DAWN)\label{aff98}
\and
Niels Bohr Institute, University of Copenhagen, Jagtvej 128, 2200 Copenhagen, Denmark\label{aff99}
\and
Universidad Polit\'ecnica de Cartagena, Departamento de Electr\'onica y Tecnolog\'ia de Computadoras,  Plaza del Hospital 1, 30202 Cartagena, Spain\label{aff100}
\and
Caltech/IPAC, 1200 E. California Blvd., Pasadena, CA 91125, USA\label{aff101}
\and
Aurora Technology for European Space Agency (ESA), Camino bajo del Castillo, s/n, Urbanizacion Villafranca del Castillo, Villanueva de la Ca\~nada, 28692 Madrid, Spain\label{aff102}
\and
Institut d'Astrophysique de Paris, 98bis Boulevard Arago, 75014, Paris, France\label{aff103}
\and
ICL, Junia, Universit\'e Catholique de Lille, LITL, 59000 Lille, France\label{aff104}
\and
California Institute of Technology, 1200 E California Blvd, Pasadena, CA 91125, USA\label{aff105}
\and
Department of Physics and Astronomy, University of British Columbia, Vancouver, BC V6T 1Z1, Canada\label{aff106}}

\abstract{The Euclid Ecliptic Survey was conducted during the calibration phase of the mission, from 23 to 31 December 2023, as a campaign to study Solar System objects. We used data from this survey to analyse more than 23\,000 appeareances of 2321 known asteroids. Due to their high apparent angular motion relative to the background stars (5--$60^{\prime\prime}\,\mathrm{h}^{-1}$), these objects appear as streaks in VIS long-exposure images. We set out to estimate their spin periods, since only $7\%$ of them have periods published in the literature. We used multiple apertures along each streak to increase the time resolution of our light curves. Our method combines a Lomb--Scargle approach with a Markov chain Monte Carlo (MCMC) algorithm to characterise the posterior distributions. Some asteroids show multimodality in the MCMC search, indicating period aliases; in these cases, we report all aliases and their likelihoods. We validate our pipeline by comparing our fitted periods with 48 published periods, including period harmonics. We find that $44\%$ of our periods are within $1\%$ of those published and $98\%$ are within $15\%$, and we establish that with $98\%$ confidence the best solution can be found among the first three aliases. All reliable periods reported are in agreement with our current understanding of the spin-period distribution for asteroids. We find 16 periods below the spin barrier of 2.2\,h with absolute magnitudes below 19, and thus 16 candidate super-fast rotators. We provide light curves for all 2321 objects observed and 889 high-quality periods in an open-access catalogue. The asteroids with reported periods include five Mars crossers, four Cybeles, four Hildas, three Hungarias, and 877 asteroids in other regions of the main belt. Our results represent the first batch of spin periods extracted from \Euclid light curves and include the first-ever period measurements for $93\%$ of the objects.}

\keywords{Minor planets, asteroids: general -- Surveys -- Astronomical data bases}

   \titlerunning{\Euclid\/: Asteroid rotation periods from the Euclid Ecliptic Survey}
   \authorrunning{B.\,Y.~Irureta-Goyena et al.}
   
   \maketitle
   \nolinenumbers

\section{\label{sc:Intro}Introduction}

Time-resolved photometry can reveal key asteroid properties. For principal-axis rotators, rotation appears as a periodic signature in the asteroid light curve that can be used to determine the rotation period, also known as the spin period. In turn, the spin period can help with the analysis of other physical characteristics. It can constrain asteroid surface features, since surface smoothness relates to faster spin rates \citep{persson2022stability}, as well as the internal structure, since fast rotators are thought to be monolithic, while rubble-pile asteroids with relatively short periods may suffer from rotational instabilities leading to failure modes \citep{pravec2002asteroid, hirabayashi2015internal}. The spin period can help us study the thermophysical properties and grain size of the observed asteroids \citep{Spencer1989, lagerros1996thermal,muller1998asteroids,rozitis2011directional,marciniak2019thermal}. This is relevant for interpreting the thermal infrared emissions using telescopes such as the Wide-field Infrared Survey Explorer \citep{wright2010wide} or the upcoming European Space Agency's Near-Earth Object Mission in the Infra-Red \citep{conversi2024neomir} and NASA's Near-Earth Object Surveyor \citep{mainzer2023near}. The shape of an asteroid is influenced by its current and past spin rates \citep{barnouin2019shape}, and through shape models we can connect the spin period to the mass distribution of the asteroid \citep{ostro1995radar} and to the features of any binary companions \citep{pravec2006photometric, scheeres2006dynamical}. The sources of asteroid rotation, mainly past collisions and the Yarkovsky--O’Keefe--Radzievskii--Paddack (YORP) effect, can also be studied using spin-period analysis \citep{Rubincam2000,pravec2002asteroid,taylor2007spin, marzari2020evolution}. 

While these findings are informative at the individual level, collecting a statistically significant sample of asteroid spin periods can provide insights into the formation history of our Solar System. In particular, the asteroid main belt (MB), located between Mars and Jupiter, is relevant to understanding aspects such as the migration of the giant planets to the outer Solar System \citep[and references cited therein]{clement2020record} or the origin of near-Earth objects \citep{granvik2017escape}. However, as of November 2025, periods have been published for fewer than 5\% (around 50\,000) of the known asteroids \citep{berthier2023ssodnet}. Of these, only about 20\% of the periods that include reliability flags are considered to be of high quality. The light curve analysis required to determine the spin period often needs many high-quality data points, and the brightness variations, if subtle, can lie below the noise level of small telescopes. 

In this work, we exploit the capabilities of the European Space Agency's \Euclid space telescope \citep{EuclidSkyOverview} to expand our catalogue of asteroid spin periods. Although large astronomical surveys covering extensive portions of the sky have produced large amounts of asteroid data in the last decade, their observing strategies have typically yielded sparse photometry for these objects. This is the case for the Zwicky Transient Facility \citep{bellm2018zwicky}, Pan-STARRS \citep{kaiser2002pan}, and the Asteroid Terrestrial-impact Last Alert System \citep{2018PASP..130f4505T}. Conversely, dense asteroid light curves that can be used to determine spin periods accurately remain relatively scarce in the literature and have never been gathered for the faintest apparent magnitudes in the range of the \Euclid Visible Camera \citep[VIS; ][]{EuclidSkyVIS}, between 20 and 24. The observing strategy of \Euclid, even if not primarily designed for asteroid science but for cosmological studies, nevertheless allows us to obtain dense photometry and to fill an unexplored gap in the asteroid population.

As is discussed by \cite{carry2018solar}, during its nominal six-year mission \Euclid will observe up to 150\,000 Solar System objects (SSOs, which includes asteroids and comets) with extraordinary photometric stability. Since the mission will reach fainter magnitudes than previous surveys, many of these asteroids will have never been observed before. To test the potential of \Euclid to find and characterise SSOs, a dedicated campaign, the Euclid Ecliptic Survey (EES), was conducted during the \Euclid phase-diversity calibration (PDC) at the beginning of the mission, before the start of the Euclid Wide Survey \citep{Scaramella-EP1}. The PDC observations require thermal stability and it was decided to use the thermal stabilisation periods for three sets of dedicated observing programmes, one of which was for SSOs. During this campaign, from 23 to 31 December 2023, VIS was pointed directly at the ecliptic plane and approximately followed the average motion of the MB. This strategy allows us to extract denser, longer light curves and observe more SSOs than the nominal Euclid Wide Survey mode, which generally avoids the ecliptic plane and hence produces sparser SSO photometry. The EES used both \Euclid instruments, VIS and the Near Infrared Spectrometer and Photometer \citep[NISP;][]{EuclidSkyNISP}. For the purposes of this study, we only use data from VIS; the NISP analyses including thermal data are presented in a separate publication that will include a photometric analysis followed by taxonomic classification based on the multiband photometry \citep[][Pöntinen et al., in prep.]{pontinen2025asteroid}. 

\begin{figure}
    \centering
    \includegraphics[width=\linewidth]{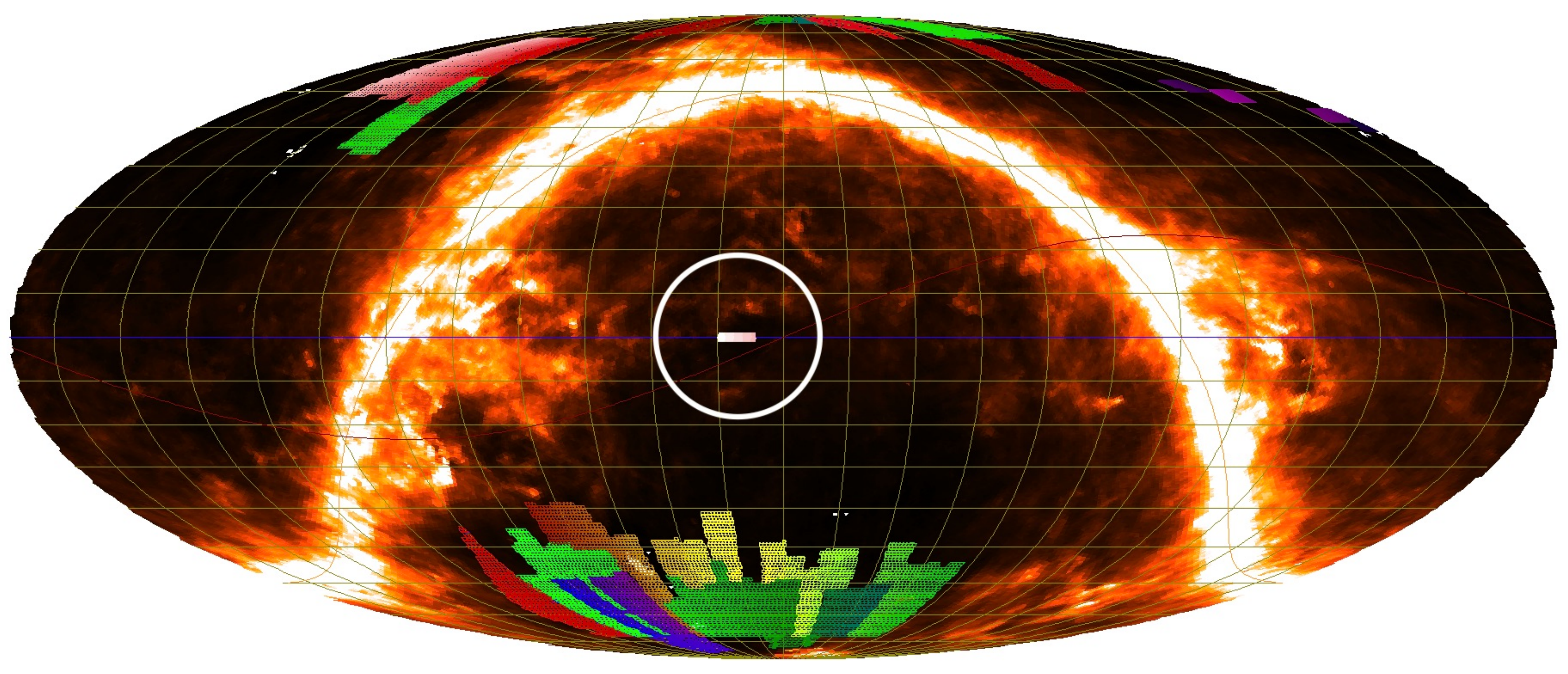}
    \caption{Position in ecliptic co-ordinates of the EES (encircled in white) in the context of the DR1 (coloured tiles), projected using the \Euclid Survey System \citep{gomez2018euclid}. The survey took place from 23 to 31 December 2023 and lay on the ecliptic plane, since it was designed to approximately follow the motion of the asteroids in the MB. A detail of the survey can be seen in Fig.~\ref{fig:eclipticsurvey}.}
    \label{fig:fullsky}
\end{figure}

\begin{figure}
    \centering
    \includegraphics[width=\linewidth]{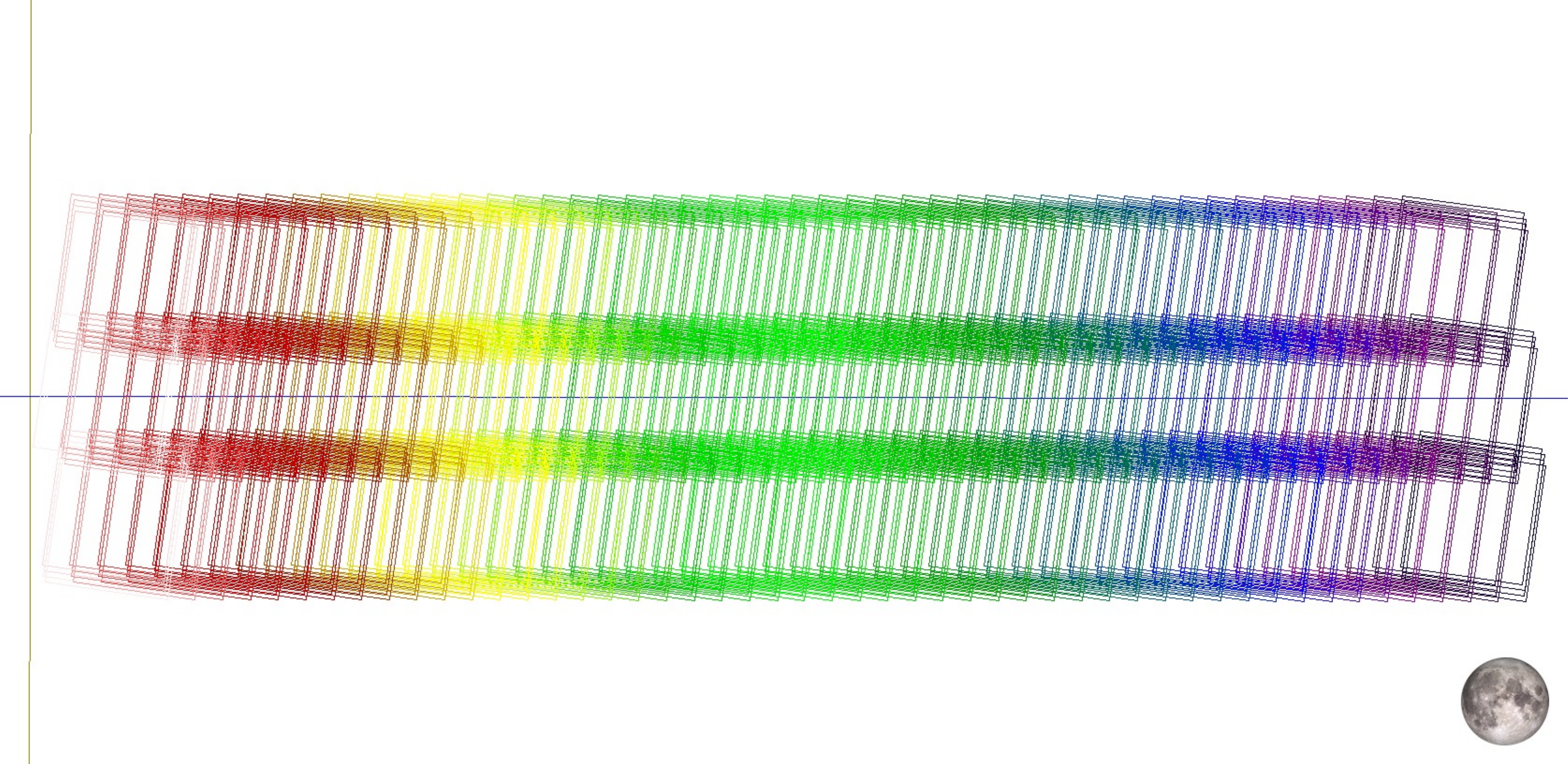}
    \caption{Detail of the observing strategy of the EES, which starts on the rightmost squares on 23 December 2023 (black) and ends on the leftmost squares on 31 December 2023 (pink). Each square corresponds to a single exposure of 560.5\,s (dither) and each observation is comprised by four dithers. Within each observation, the four dithers follow an S-shaped pattern. The observations had a high overlap to ensure that the asteroids observed would be seen in as many exposures as possible. In the figure, ecliptic longitude increases from right to left and the ecliptic latitude increases from bottom to top (ecliptic north is up). The Moon is displayed for a scale reference.}
    \label{fig:eclipticsurvey}
\end{figure}

Figure \ref{fig:fullsky} shows the EES in context, projected in ecliptic co-ordinates, and Fig. \ref{fig:eclipticsurvey} presents a detail of the observing strategy. In this work, we use all 148 observations collected during the EES. We use trailed sources detected by \cite{pontinen2025asteroid} and search for periodicity in their signal. We build upon previous research that developed semi-automated pipelines to ingest the large volume of data produced by large surveys. The work pioneered by \cite{polishook2012asteroid} and continued by \cite{waszczak2015asteroid} and \cite{chang2017asteroid} applied Fourier analysis to determine the spin periods of thousands of asteroids observed by the Palomar Transient Factory. Later studies shifted towards the use of Lomb--Scargle algorithms, which are more robust to unevenly sampled time series, as is generally the case for asteroid light curves \citep[e.g.][]{erasmus2021discovery, eglitis2025rotation}. As is discussed in these papers, the survey cadences used to obtain many of the published light curves yielded sparse photometry that could not always fully constrain the period, leading to ambiguous solutions. The newest surveys, such as the Vera C. Rubin Observatory, and the more recent use of space telescopes that provide denser light curves due to their observing strategy, such as NASA’s TESS and K2, have been shown to significantly improve previous spin-period estimates; for instance, the studies of \cite{szabo2022rotation}, \cite{sergeyev2025rotation}, \cite{vavilov2025rotation}, or \cite{Greenstreet2026}. Our work complements these efforts and paves the way for future analyses during the Euclid Wide Survey. 

The remainder of the paper is structured in five parts. Section \ref{sc:Data} describes the data products that we used for our analysis. In Sect. \ref{sec:lc-model} we develop the mathematical expressions and physical assumptions we made to construct the model for our asteroid light curves. In Sect. \ref{sc:Methods} we explain our methodology and justify any technical decisions. In Sect. \ref{sc:Results} we present and discuss the results obtained during the validation of our pipeline and include tables with our fitted periods. Lastly, in Sect. \ref{sc:Conclusion}, we lay out the main conclusions of our study and comment on future work. 

\section{\label{sc:Data}Data}

Our work focuses on asteroids moving fast relative to the background stars, with apparent motions of 5--60$\arcsecond\,\mathrm{h}^{-1}$. Our detection capabilities impose this dynamical constraint: we do not observe faster objects, which would be near-Earth objects and are much less frequent (about $1000$ times), and slower objects cannot be detected by the detection pipeline \citep[see][]{pontinen2020euclid}. The slow objects will be detected by a separate pipeline \citep{nucita2025euclid}. Since the telescope did not explicitly track these asteroids, they appear as streaks of light spanning 10--100 pixels in the VIS long exposures of 560.5\,s (the pixel size of VIS is $\ang{;;0.1}$\,pixel$^{-1}$). This allows us to extract multiple data points from each exposure and construct dense light curves to determine the spin period accurately. For our analyses, we used calibrated (Level 2) images retrieved directly from the \Euclid data products in the \Euclid Science Archive System, Observation IDs 66202--66350, with a stable zero-point magnitude of around 24.4. The VIS observation range is in the optical band (550--900\,nm) and the instrument does not use any filters. A thorough description of the calibration and processing of these images can be found in \cite{Q1-TP002}. All data products used will be published as part of the Euclid Data Release 1 \citep[DR1; ][]{DR1cite}.

We used the catalogue of approximately 23\,000 streaks detected by \cite{pontinen2025asteroid}, which provides the equatorial (RA, Dec) endpoint co-ordinates. The streaks were found using \texttt{StreakDet}, a streak-detection software originally developed by \cite{virtanen2016streak} to find fast-moving objects in telescopic surveys of space debris. The \texttt{StreakDet} configuration parameters were optimised using simulated \Euclid VIS images \citep{pontinen2020euclid}. The software was first applied to all VIS individual long exposures, producing output streak catalogues with a high number of false positives, caused by other linear objects in the image, mainly diffraction spikes. The VIS short exposures were not used by the detection pipeline. A post-processing step using interdither linking was introduced to identify streaks moving consistently between exposures, requiring them to appear in at least three exposures to be valid. This stage decreased the percentage of false positives to around $1\%$. According to the tests reported by \cite{pontinen2020euclid}, \texttt{StreakDet} was estimated to achieve its best recovery results when applied to streaks brighter than magnitude 23 in the optical band and faster than $8\mathrm{\arcsecond\,h^{-1}}$. Following the interdither linking, we relate them to known objects using the tool of the Virtual Observatory service \texttt{SkyBot} \citep{berthier2006skybot}, with an allowed uncertainty of $1^{\prime\prime}$. This allows us to link the streaks further across observations and considerably extends the observation baseline, resulting in each of the 2321 objects being observed an average of 10 times. Without this step and given the length of the \Euclid observations, the period determination would not be possible, except for very short periods (below 1\,h). For this reason, in this paper we focus on analysing the streaks of known objects and we leave the streak-linking and analysis of unknown objects found by \cite{pontinen2025asteroid} for future work. For a detailed account of the streak catalogue, see \cite{pontinen2025asteroid}.  \\

\section{Light curve model}
\label{sec:lc-model}

The apparent brightness of a rotating asteroid is affected by its shape, solar phase angle, and the distances to the Sun and the observer. The variation due to rotation is dominated by the shape of the object; albedo variations can also contribute, but for objects much larger than our target population \citep{magnusson1991albedo, wiktorowicz2017albedo}. We followed the approach laid out in \cite{harris1989photoelectric}, but  we treated the solar phase angle and viewing geometry of each object as constant, since, in our case, the observations only cover a time span of several hours or, at most, several days. We can fit a model of the asteroid flux $\Phi(t; \alpha)$ to the resultant light curve using a truncated Fourier series of $n$ harmonics,
\begin{equation}
\Phi(t; \alpha) =
\sum_{k=1}^{n} A_k
\sin\!\left(\frac{2\pi k t}{P} + \phi_k\right) + C\,,
\label{eq:fourier}
\end{equation}
where $P$ is the period, $A_k$ and $\phi_k$ are the amplitude and phase of the $k$th harmonic, and $C$ is the average brightness. The parameter vector $\alpha$ includes a term $\ln S$, the Gaussian intrinsic scatter \citep[for a detailed discussion, see][]{Kelly2007}, which is capped at the maximum photometric uncertainty and accounts for additional variability or other sources of uncertainty not modelled, such that $\alpha = (A, P, \phi, C, \ln S)$. 

For simplicity, in our analysis we retain the single-harmonic representation ($n=1$), which is proved to be sufficient later in our study. Even if the asteroids are not perfectly ellipsoidal, we assume that they present two maxima and two minima per rotation (double peak), so that we can fit a sine period and interpret it as half of a rotation period,
\begin{equation}
\Phi(t; \alpha) =
A \sin\!\left(\frac{2\pi t}{P} + \phi\right) + C\,,
\label{eq:single-sine}
\end{equation}
where the period of the asteroid, $P_\mathrm{ast}$, is $P_\mathrm{ast}=2P$.

We denote the observation times as $T = \{t_j\}$ and the corresponding photometric uncertainties as $E = \{\sigma_j\}$, where we assume that the uncertainty of each measurement with respect to the model is independent and Gaussian. This allows us to express the likelihood over the entire light curve $Y = \{y_j\}$,
\begin{equation} 
\mathcal{L}(Y \mid T, E, \alpha) = \prod_{j=1}^{N} \frac{1}{\sqrt{2\pi(\sigma_j^2 + S^2)}} \exp\!\left[ -\frac{(y_j - \Phi_j)^2}{2(\sigma_j^2 + S^2)} \right]\,, 
\label{eq:likelihood} 
\end{equation}
where $\Phi_j = \Phi(t_j; \alpha)$ is the model at time $t_j$. 

To further constrain the parameters of our model, we adopted priors with hard parameter bounds. This ensured that all parameters remained within a physically meaningful range of values. We can note the prior distributions as $p(\alpha \mid \psi)$, where $\psi$ conveys the set of prior bounds. Using Bayes' theorem and Eq. \eqref{eq:likelihood}, the posterior distribution is then given by 
\begin{equation}
p(\alpha \mid T, Y, E, \psi)
\propto
\mathcal{L}(Y \mid T, E, \alpha)\, p(\alpha \mid \psi)\,,
\label{eq:posterior}
\end{equation}
which serves as the basis for the optimisation and sampling methods that are described in the following section. 

\section{Methods}
\label{sc:Methods}
\begin{figure*}
    \centering
    \includegraphics[width=\linewidth]{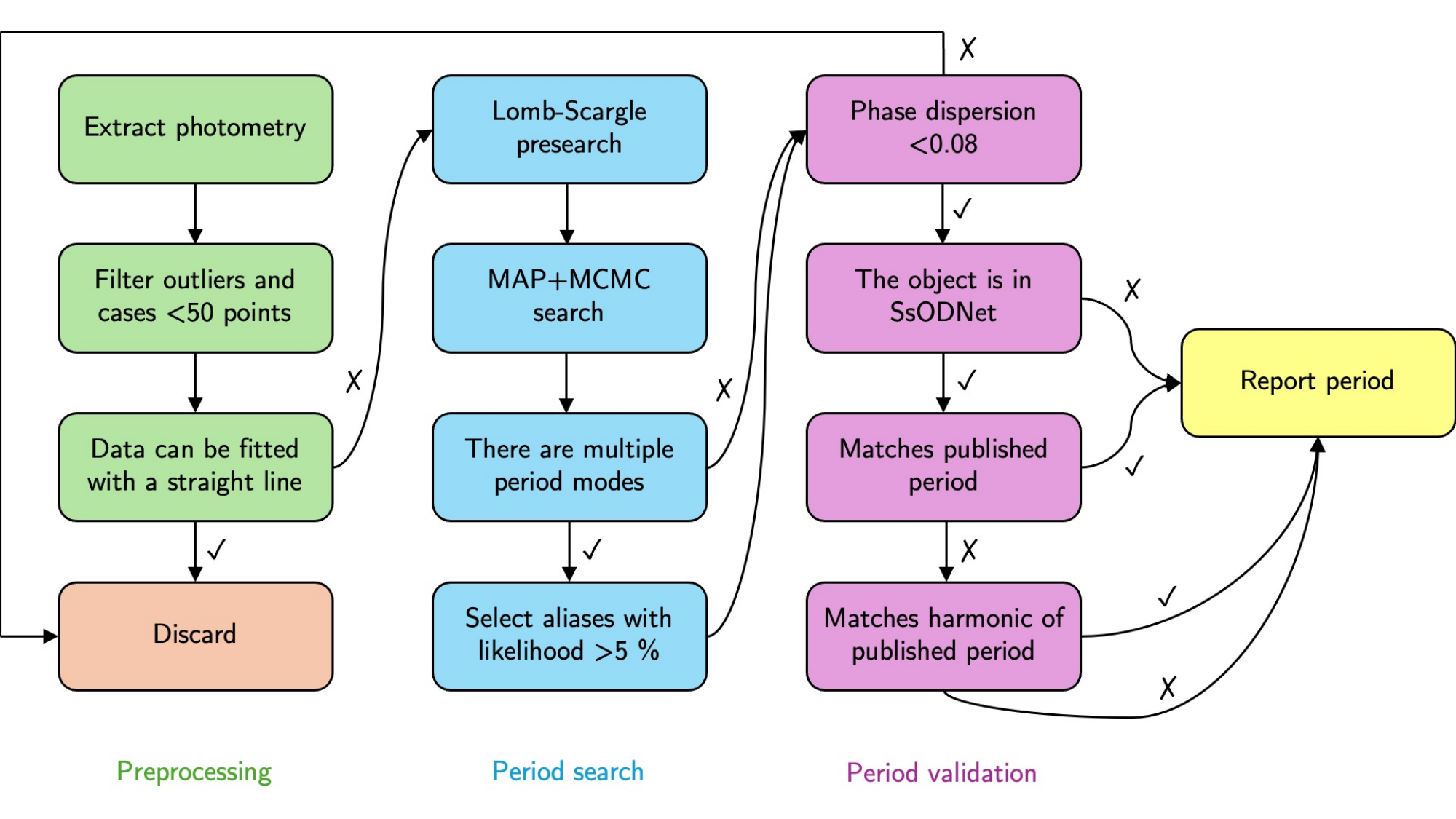}
    \caption{Diagram summarising the general workflow of our pipeline. We first build our light curves, excluding the outliers and cases with insufficient signal. Then, we search the period using multiple algorithms and, after filtering, compare our results with those published in the catalogue.}
    \label{fig:flowchart}
\end{figure*}

The methodology described in this section is summarised in Fig. \ref{fig:flowchart}. Our approach is divided into three parts: pre-processing, which includes extracting the photometry and filtering the outliers, period finding, which includes several search algorithms, and period validation, which uses two different tests to assess the validity of the results. 

\subsection{Photometry extraction}
\label{subsection:photometry}

Traditionally, asteroid-tracking surveys have used circular aperture photometry on their targets, which appear as point sources amidst a background of trailed stars. On the contrary, in this work the stars are point sources, while the asteroids are streaked. This approach offers the advantage of a higher sampling rate than a point source but challenges conventional photometric approaches. To overcome this challenge, we highlight two existing solutions: \texttt{TRIPPy}, by \cite{fraser2016trippy}, and the more recent method described in \cite{devogele2024aperture}. We chose the latter approach, since it is better optimised to extract the photometric information needed to constrain the spin period. 

\begin{figure}
    \centering
    \includegraphics[width=\linewidth]{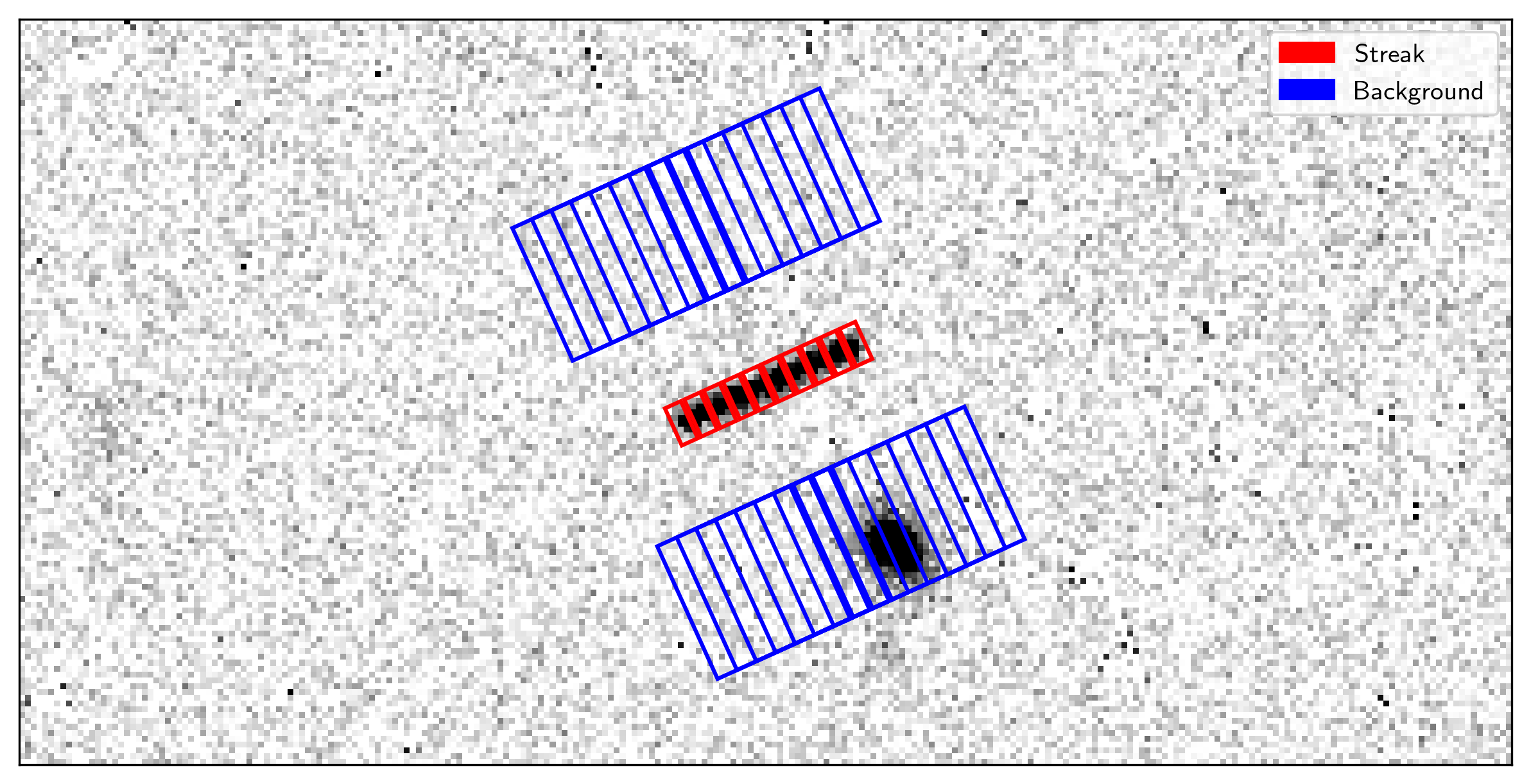}
    \caption{Example streak with apertures placed along the direction of motion of the asteroid to increase the temporal resolution of the sampling. A bright galaxy falls on several of the background apertures; the approach to address it is described in Sect.~\ref{subsection:photometry}.}
    \label{fig:apertures}
\end{figure}

In this approach, we covered the streak with rectangular apertures instead of the conventional circular apertures used for point-source objects. This allowed us to align several apertures along the direction of the streak, each covering a small fraction of it, and to maximise the signal-to-noise ratio (S/N) collected by each aperture. We placed two additional larger apertures to compute the level of the sky background, one on each side of each streak aperture and at a sufficient distance (10\,pixels) to avoid collecting any signal from it. The distribution of the apertures taken can be seen in Fig.~\ref{fig:apertures}. 

The streak aperture size was determined by considering the following four conflicting objectives. 
\begin{itemize}
    \item Maximise the temporal resolution, which requires the direction along the streak (length) to be as short as possible.
    \item  Maximise the S/N by increasing the aperture length.
    \item Minimise the flux loss, which requires the direction perpendicular to the streak (width) to be as large as possible to encompass the streak fully. 
    \item Maximise the S/N, which requires the width not to exceed the streak bounds.
\end{itemize}
The four criteria are balanced empirically for the target \Euclid asteroids. We find that an aperture size of 5~pixels along the streak~$\times$~7~pixels perpendicular to it guarantees a high temporal resolution, even below 1\,min, while optimising the S/N. This choice is conservative in width to account for bright asteroids, which extend over a larger number of pixels. For each aperture, the uncertainty in the measured flux is calculated using the CCD equation, which accounts for flux, background noise, and the aperture sizes of both the streak and the background. In this case, the dark current and readout noise are negligible. A more detailed discussion of these considerations can be found in \cite{devogele2024aperture}. 

\begin{figure}
    \centering
    \includegraphics[width=\linewidth]{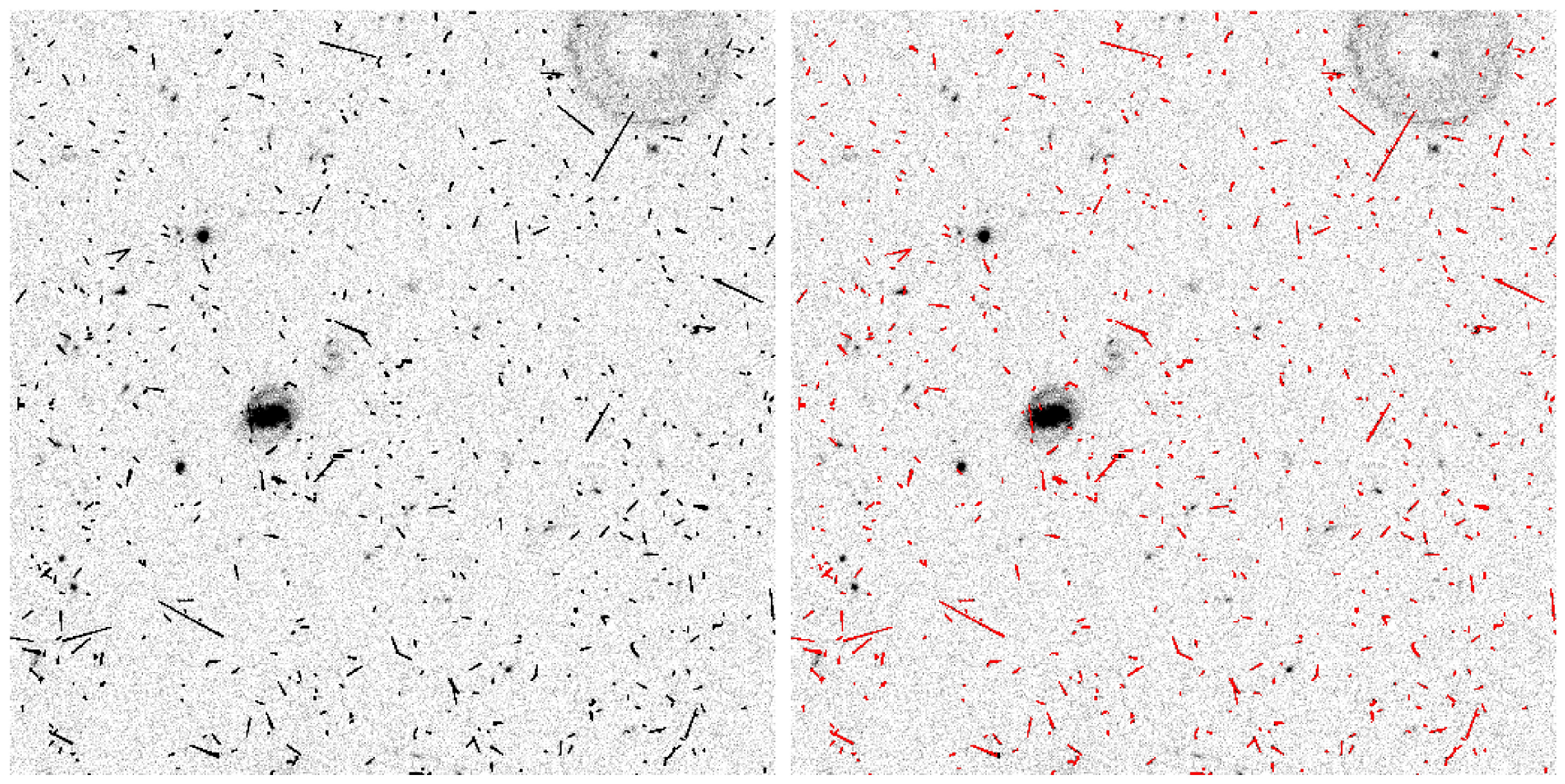}
    \caption{Cut-out of VIS image comprising (left) and the same image with the overlaid mask of cosmic rays in red (right).}
    \label{fig:cosmicmask}
\end{figure}

Since \Euclid is located outside of Earth's atmosphere, the images contain a large number of cosmic rays ($10^5$ per exposure during the EES, a level that can increase significantly during periods of intense solar activity). During the photometric reduction of the Level 2 data, the Euclid Consortium pipeline incorporated flags that recorded the exact positions of all cosmic rays in each exposure. However, we observe that these flags, which are designed for a more general cosmic-ray masking, search for isolated linear features and do not discriminate between the unwanted cosmic rays and our target asteroid streaks. For this reason, we used the cosmic-ray masks independently produced with \texttt{Astro-SCRAPPY} \citep{mccully2018astroscrappy} as a part of the pipeline built by \cite{nucita2025euclid}, specifically developed to avoid flagging asteroid streaks. We find the method to be successful at this task and highly effective at flagging cosmic rays. An example of a cosmic-ray mask that we used can be seen in Fig.~\ref{fig:cosmicmask}.

Pixels flagged as cosmic rays that fall inside one of the apertures were handled differently depending on the aperture type. In the background apertures, we removed all pixels containing cosmic rays. This removal does not bias our calculation of the sky background, which is based on the median of each background aperture. For streak apertures, we attempted statistical inference and deep-learning algorithms, but they both produced artificial flux variations unrelated to the rotation of the asteroid. Since only removing the affected pixels would lower the flux extracted from that aperture, artificially creating an unphysical drop in the light curve, we decided to remove the entire aperture in these cases. Beyond cosmic rays, apertures can also be contaminated by other objects such as stars or galaxies, as in the example in Fig. \ref{fig:apertures}. Background apertures are least affected by this type of unwanted flux since we use sigma clipping and then take the median of the resultant values. Streak apertures affected are removed using a more complex outlier filtering at a later stage.

\begin{figure}
    \centering
    \includegraphics[width=\linewidth]{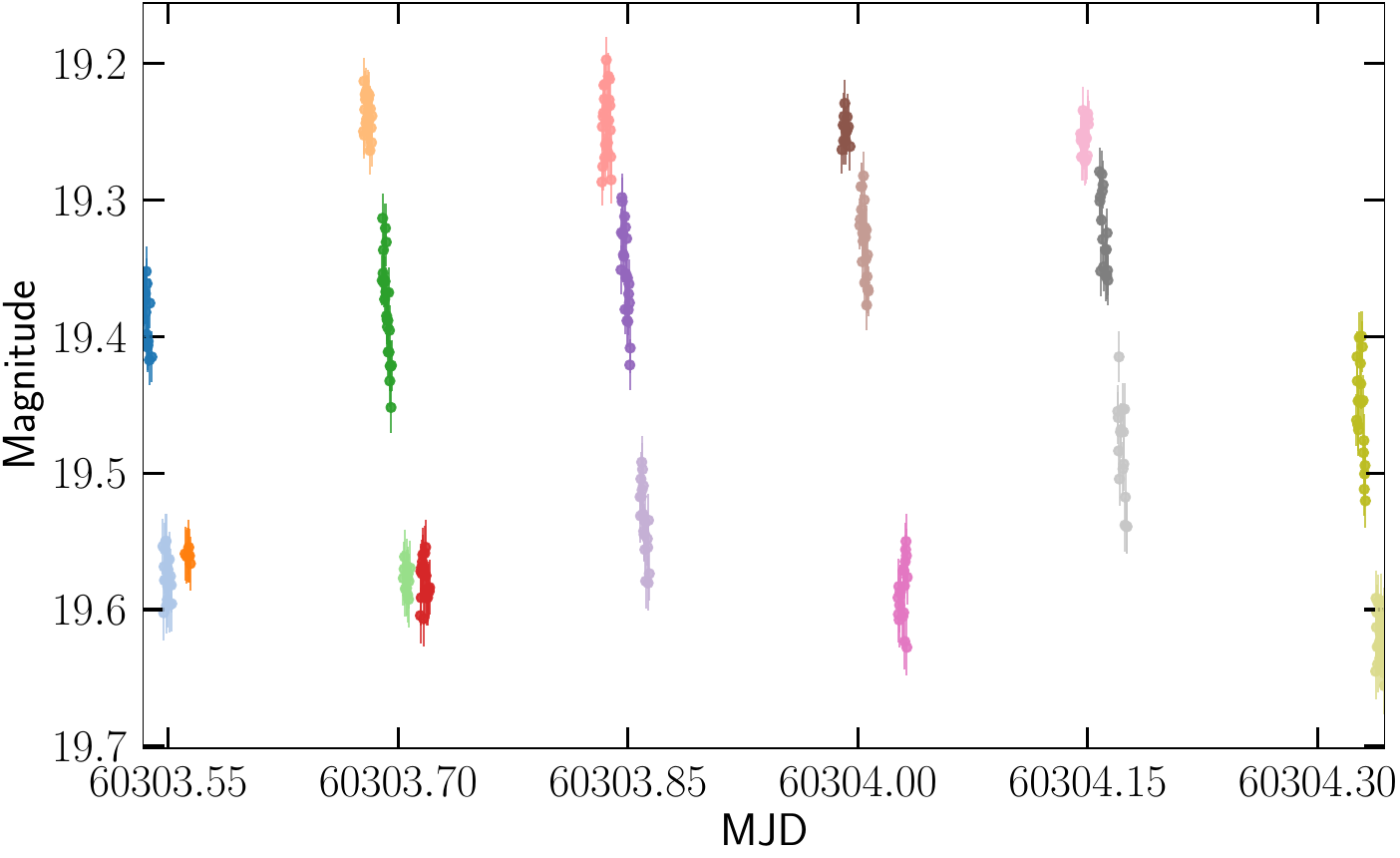}
    \caption{Light curve of asteroid (18957) Mijacobsen. The time range covers several rotations; however, the sampling intervals are highly regular, which can potentially cause period aliases (a possibility discussed in Sect.~\ref{subsubsec:aliases}). Each data cluster and colour corresponds to a different streak of origin.}
    \label{fig:mijacobsen}
\end{figure}

\begin{figure}
    \centering
    \includegraphics[width=\linewidth]{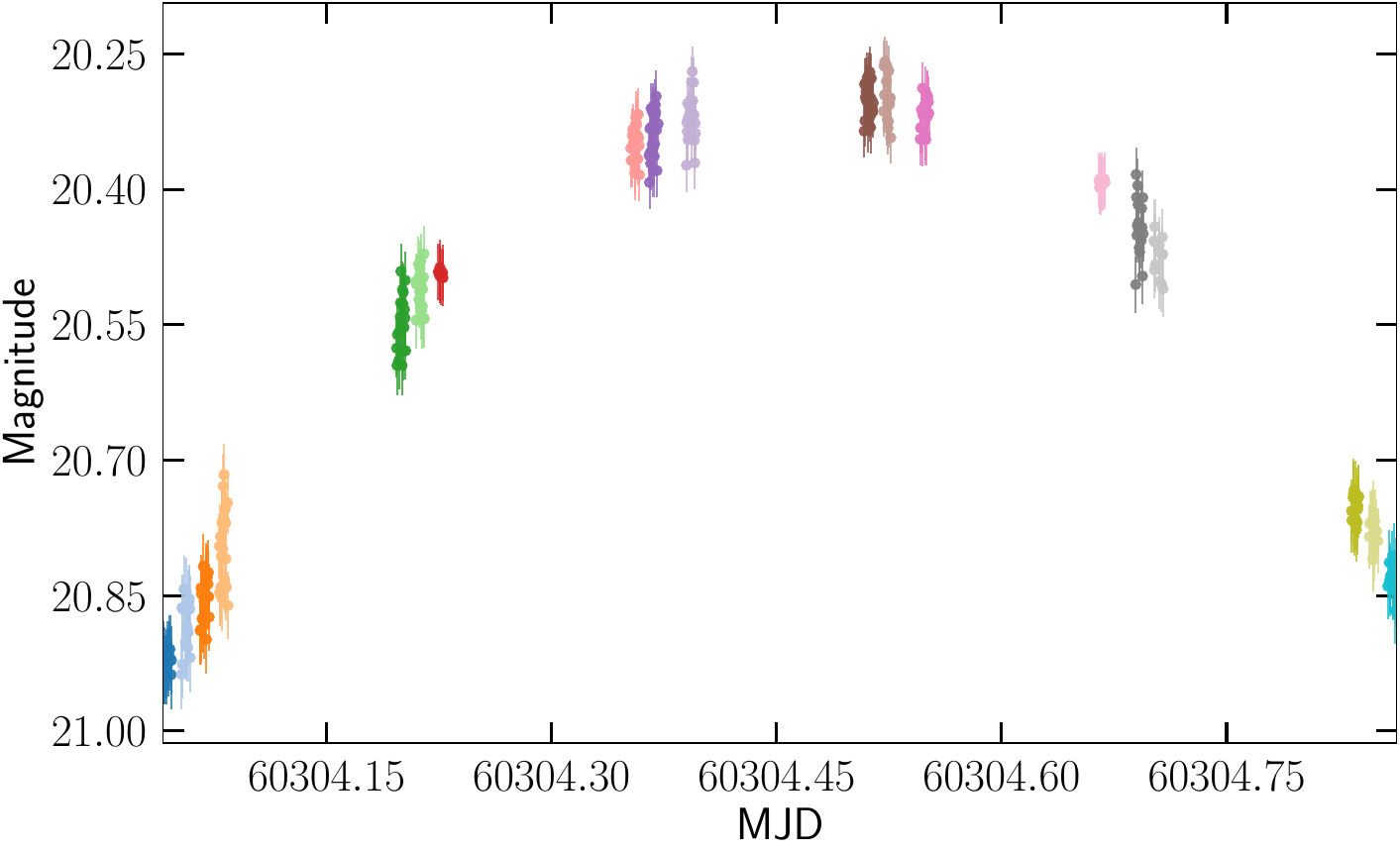}
    \caption{Light curve of asteroid 2000 SQ$_{24}$. The time range covers less than one full rotation, which makes the spin-period determination highly suspect. Each data cluster and colour corresponds to a different streak of origin.}
    \label{fig:2000_SQ24}
\end{figure}

\begin{figure}
    \centering
    \includegraphics[width=\linewidth]{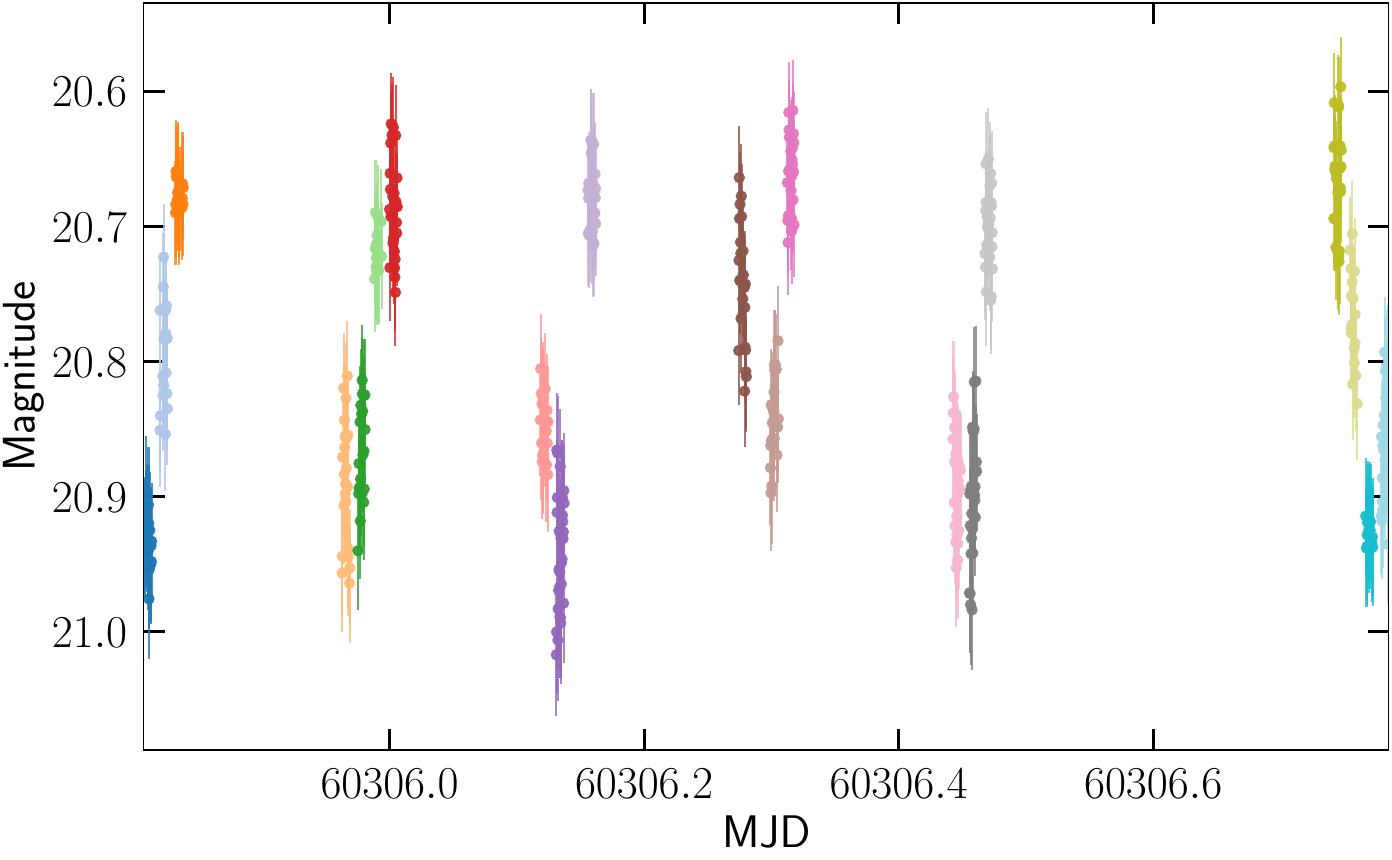}
    \caption{Light curve of asteroid 2005 RE$_{29}$. The time range covers several rotations, which allows for reliable spin-period determination. Each data cluster and colour corresponds to a different streak of origin.}
    \label{fig:2005_RE29}
\end{figure}

To build our light curves, we determine the timestamp of each aperture by mapping the endpoints of the streaks to the start and end times in the header (expressed in modified Julian date, MJD) and dividing the total exposure time by the number of apertures. We verified the direction of the streaks by confirming that the sign of the RA, Dec variation within each streak matched the sign of the RA, Dec variation between streaks. To improve the spin-period analysis, we post-processed the light curves, trimming the first and last points, which have shorter exposure times, and removing the outliers. This ensures that the period determination is not artificially affected by contaminating objects covering the streaks, such as galaxies, stars, or image artefacts. We flag outliers with a winsorised sigma-clipping algorithm that removes points more than $2\,\sigma$ away from the median of each streak or $2\,\sigma$ away from the median of each light curve. This routine is iterated until convergence and allows for the subsequent determination of the period. Our tests show that without the sigma clipping, the period fit is wrongly driven by the outliers. We show several examples of our light curves in Figs.~\ref{fig:mijacobsen}--\ref{fig:2005_RE29}.

\subsection{Period determination}
\label{subsection:period_determination}
\subsubsection{Preliminary vetting}
\begin{figure}
    \centering
    \includegraphics[width=\linewidth]{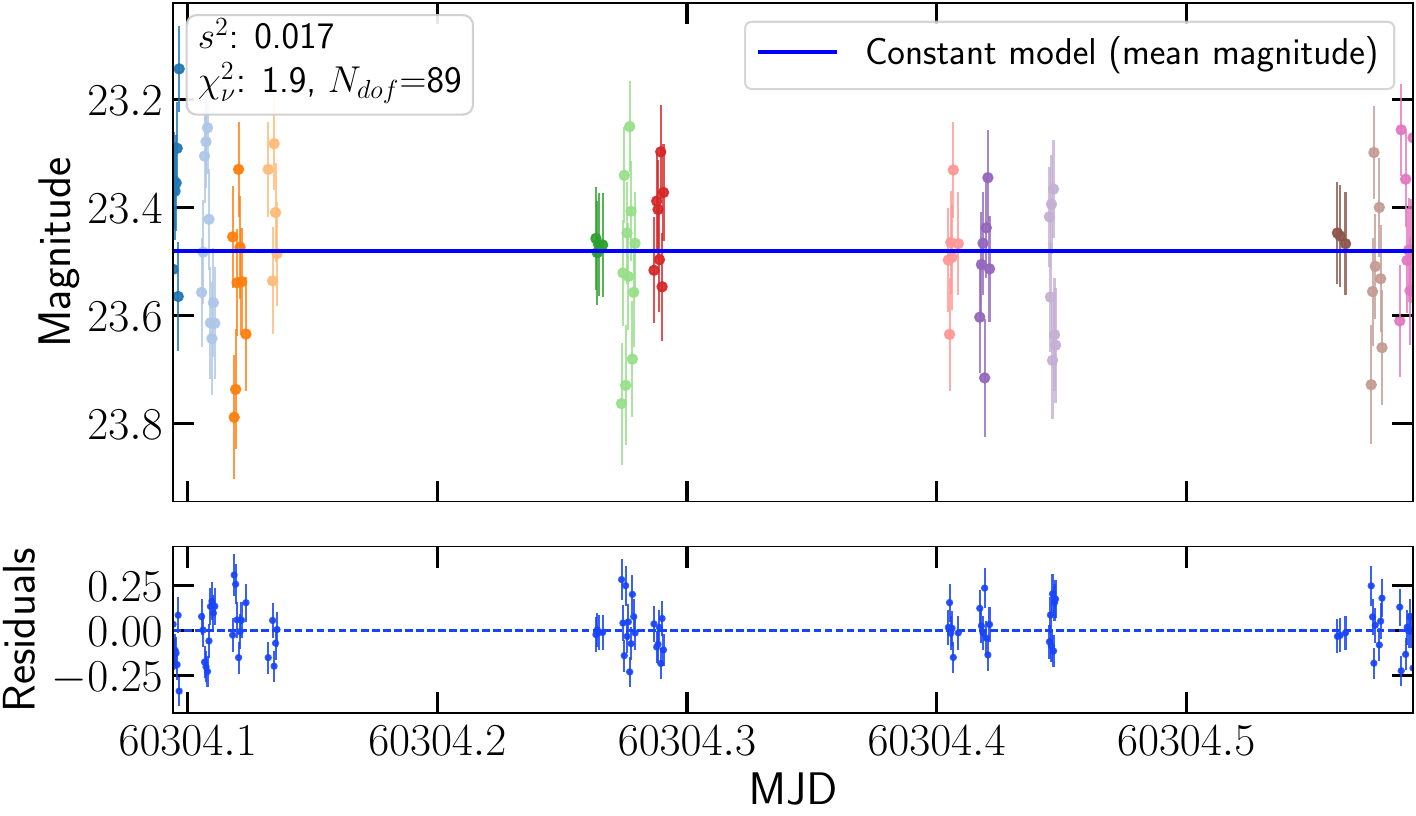}
    \caption{Light curve of asteroid 2010 ES$_{89}$ fitted by a constant. The data are coloured by streak of origin. The fit is adequate, which implies that the data do not have enough signal to allow for a high-quality period search. We discarded this light curve. }
    \label{fig:constant}
\end{figure}

In certain cases, either because the coverage is too short or because the amplitudes of the light curves are smaller than the measurement uncertainties, the light curves do not contain enough signal to constrain a period. As is shown in Fig. \ref{fig:constant}, we flag these cases by fitting a constant through each light curve; whenever the reduced chi-square, $\chi^2_\nu$, is 2 or below, we discard the light curve for period analysis. We also discarded light curves with temporal coverage below 1\,h after observing that in these rare cases, the algorithm always fits the noise instead of the data. The discarded light curves are presented in Appendix~\ref{sc:appendix_discarded}. We find that using this conservative threshold flags cases that lack enough variability for the period search. We then searched for rotation signatures in the light curves using a three-stage approach: Lomb--Scargle periodogram for initial period estimation, maximum a posteriori (MAP) optimisation to refine the estimate, and Bayesian inference with Markov chain Monte Carlo (MCMC) to find the best value. 

\subsubsection{Lomb--Scargle pre-search}
As is discussed in Sect. \ref{sc:Intro}, the Lomb--Scargle algorithm is more robust than simple Fourier analysis for unevenly sampled time series \citep{vanderplas2018understanding}, which is generally the case for asteroid light curves. It is often applied to infer a period that is used as an initial guess when optimising the model or refined using other search methods, such as MCMC. In the case of the \Euclid asteroids, the number of data points, the number of observations, and the data sparsity can vary from one object to another, which makes the use of the Lomb--Scargle method preferable. We used the \texttt{nifty-ls} method developed by \cite{garrison2024nifty}, which is faster and more accurate than the default Astropy implementation. We used the power spectrum from the Lomb--Scargle estimation to set the bounds for the subsequent search, which ranges from 0.5--2.0 times the Lomb--Scargle period. 

\subsubsection{Bayesian inference via MCMC}
Following the approach discussed in \cite{foreman2013emcee}, we improved the parameter estimates before the MCMC run by optimising the posterior in two stages. We first explored the parameter space with a global search using differential evolution within the bounds of the prior. Then, the best candidates were locally refined using an L-BFGS-B optimiser \citep{byrd1995limited}. The result of this MAP estimate was used to initialise the MCMC walkers. This step does not impose any bounds on the search, but we observe it improves the convergence of the MCMC search. 

After that preliminary optimisation had provided sensible starting values for the model parameters, we explored their full posterior distribution using Bayesian inference and the equations laid out in Sect. \ref{sec:lc-model}. This approach allowed us to quantify parameter uncertainties and correlations, and to assess the quality of the final model. To build our chains, we used the affine-invariant ensemble sampler implemented in the \texttt{emcee} package developed in \cite{foreman2013emcee}. In practice, each MCMC run was initialised near the parameter values found by the MAP optimiser, with small random perturbations to ensure there is diversity among the walkers. We followed the recommendations of the \texttt{emcee} documentation, which suggests using move mixtures to favour transitions between modes in multimodal posteriors. The lower bound for the period was 0.02~days (approximately $30$\,min) to include possible fast rotators. Faster periods are possible \citep[see][]{Greenstreet2026}, but we expect them only very rarely in our target population, and we observe that reducing the lower bound causes the algorithm to fit the noise. The upper bound was set individually for each asteroid at twice the coverage of the observations; it should be noted that for periods found to be greater than the coverage, the result must be taken with caution and follow-up observations are required. We used 16~walkers, a number that exceeds twice the number of dimensions fitted, which is a requirement laid out in the \texttt{emcee} documentation. The samples of all jointly evaluated parameters represent the posterior distribution. We evolved the walkers for 100\,000 steps, which is at least an order of magnitude greater than the estimated integrated autocorrelation time of the chains to ensure stable posterior estimates. We observe that increasing the number of steps beyond that value does not significantly affect the MCMC results, indicating that the chains converge well with the chosen value. 

\subsubsection{Best period and aliases}
\label{subsubsec:aliases}

Particularly for light curves with large gaps or few data points, the spin-period solution can be degenerate, and the MCMC chain can have multiple modes, known as aliases. This is illustrated in Fig. \ref{fig:overlayedreynolds}. In such multimodal cases, our data do not fully constrain the model, leaving multiple possible solutions for the period. To account for this, we take the median of the samples assigned to the most likely mode and report those parameters as the best fit. We also report all modes containing more than 5\% of the samples as period aliases. The uncertainty of each mode, 1\,$\sigma$, is taken as the 16th--84th percentile range of the samples assigned to that mode. 

\subsection{Performance assessment}
\label{subsection:performance}
We first evaluated the quality of our estimates using a phase-dispersion metric. The phase dispersion can quantify whether the data cluster adequately when folded at a given period without assuming a specific model. We folded the light curves by the best period such that we map all points onto the phase interval [0, 1). Grouping the points in bins of ten points, we computed the average of the variances inside each bin, $\sigma^2$, and compared it to the overall variance of the folded light curve, $s^2$. Considering that we use bins with equal number of points, our definition of the phase dispersion, $\Theta$, is equivalent to that laid out in \cite{stellingwerf1978period},
\begin{equation}
    \Theta = \frac{\sigma^2}{s^2}\,.
\end{equation}
A value of $\Theta$ close to 0 indicates an extraordinary good fit, whereas $\Theta$ close to 1 suggests a poor period estimate. We establish a threshold of $\Theta=0.8$, above which we observe that the phase-folded light curves do not have a meaningful structure and discard them (although we still publish them in Appendix~\ref{sc:appendix_discarded}). We report the $\Theta$ associated with the remaining periods to express, in each case, the reliability of our estimate. In addition, we report the phase coverage, $\phi_{\rm c}$, which quantifies what fraction of the rotation is sampled by our data. To calculate it, we divide the phase-folded light curve in ten equal bins and compute how many contain data points. In cases where $\phi_{\rm c}$ is below 0.5, the fitted period should be interpreted with caution.

\section{Results and discussion}
\label{sc:Results}
\subsection{Pipeline validation}
\label{subsection:Validation}
We validated the performance of our pipeline by comparing our period estimates and their aliases with those published in the literature. We relied on real data to validate our method since there are no simulated datasets containing asteroid periods for \Euclid. We chose to use the comprehensive Virtual Observatory web service Solar system Open Database Network (SsODNet) because it offers free access to a compilation of more than 3000 articles constantly being updated (see Appendix~\ref{sc:used_periods}), including millions of SSO parameters \citep{berthier2023ssodnet}. We accessed the server via the \texttt{rocks} Python client.\footnote{\texttt{https://github.com/maxmahlke/rocks}} Whenever an object has multiple peer-reviewed periods in SsODNet, we selected the closest match. Since the period quality flag is only available for a subset of the periods, we did not discard objects that had a low-quality flag. Only 60 of the periods in our final catalogue have been published by any of the sources in SsODNet, and we do not attempt a comparison with the periods that are more than twice our coverage using \Euclid, which reduces the sample to 48 periods, all shown in Table \ref{tab:tableknown}. We include the average magnitude, $\langle m \rangle$, which is calculated by converting $C$ into magnitude, and the peak-to-peak amplitude, $A_{\rm p-p}$, calculated using $C$ and $A$. In cases in which our $A$ is strongly overestimated and the calculation fails, we report a conservative $A_{\rm p-p}$ calculated as the difference in magnitude between the maximum and minimum flux. In these and other cases where the amplitude approaches 1 mag, the adopted light curve model does not reliably predict the amplitude, but this does not affect the validity of the period. None of the known periods had $\Theta$ above 0.8.

\begin{figure}
    \centering
    \includegraphics[width=\linewidth]{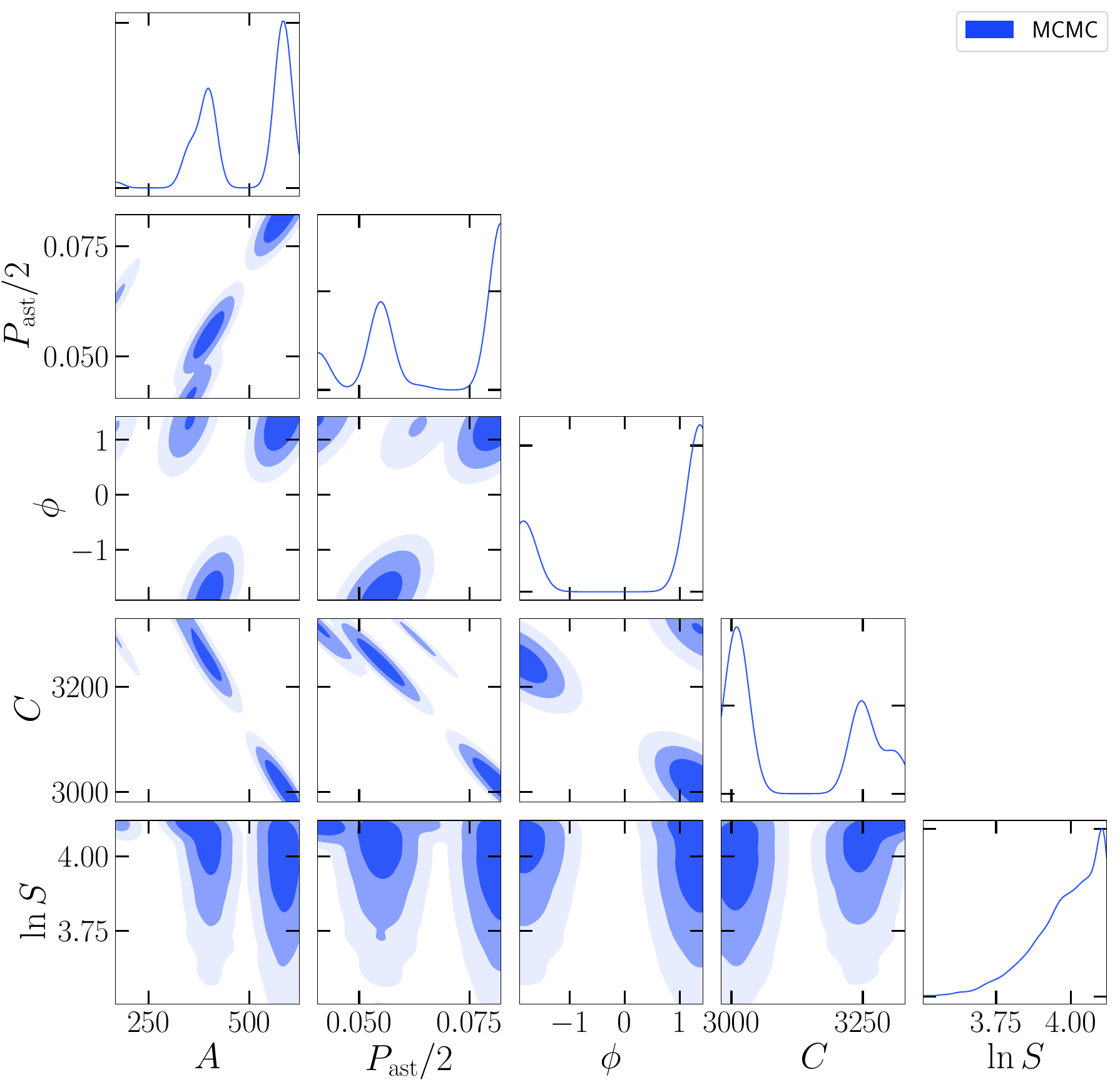}
    \caption{Posterior distributions found during an MCMC search for asteroid (12776) Reynolds. The $P_{\rm ast}/2$ panels show three clear modes for the period, which correspond to period aliases. In this case, the data have failed to constrain a unique solution. The range plotted is centred on the posterior and only a fraction of the searched space.}
    \label{fig:cornerplot}
\end{figure}

\begin{figure}
    \centering
    \includegraphics[width=\linewidth]{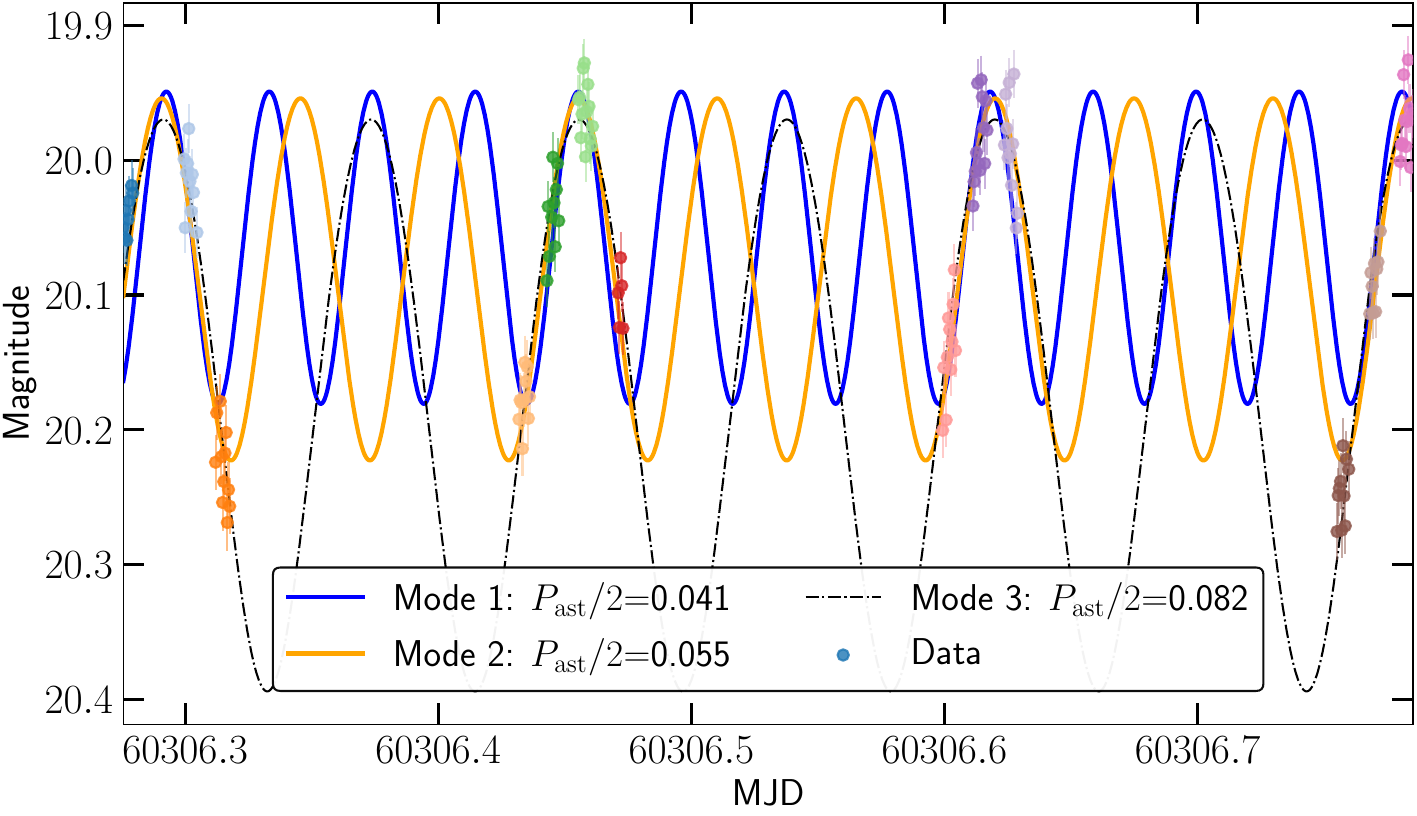}
    \caption{Overlay of the three period aliases found in an MCMC run for asteroid (12776) Reynolds. Each data cluster and colour corresponds to a different streak of origin. Mode 3 is the most likely, followed by mode 2 and mode 1. }
    \label{fig:overlayedreynolds}
\end{figure}

We compared our spin periods with all published periods in the database and the first few integer harmonics of the published periods: $2P_{\mathrm{ast}}$, $3P_{\mathrm{ast}}$, $4P_{\mathrm{ast}}$, and their inverses $P_{\mathrm{ast}}$/2, $P_{\mathrm{ast}}$/3, and $P_{\mathrm{ast}}$/4. Since harmonics convey physically equivalent descriptions of the same rotation state, the period can be reported in SsODNet with one of the harmonics, but represent an identical rotation. As opposed to aliases, which arise from sparse data, period harmonics are indistinguishable from each other; a rotation of period $P_{\mathrm{ast}}$ with a two-peak shape would present the same light curve as a rotation of period $P_{\mathrm{ast}}$/2. Of the 48 matches with the database, a quarter ($25.0\%$) of the periods have the closest match with a harmonic of the published period. For consistency, we report twice the fitted sine period, corresponding to a double-peaked light curve, although we note the harmonics and the fundamental period are mathematically equally valid solutions that cannot be distinguished solely from the photometry. 

In cases in which our periods are multimodal, we compare the published period with all aliases. In this development, we confirm that none of the parameters can provide an unequivocal identification of a specific alias as the true value. For instance, the MCMC run performed on asteroid (12776) Reynolds can converge to three different modes, 0.16~days (the published value), 0.08~days (a harmonic of the published value), and 0.06~days. The resultant posterior distribution can be seen as a corner plot in Fig. \ref{fig:cornerplot}, which shows three distinctive peaks in the period--period panel corresponding to the half period of the asteroid. Figure \ref{fig:overlayedreynolds} illustrates why the three aliases cannot be unequivocally distinguished: the large and regular gaps between the different data clusters, corresponding to different observations, allow for the presence of multiple sine peaks. The first mode allows for three peaks in each data gap, the second allows for two, and the third allows for one. The third mode of 0.16~days, the published value, is the most likely fit in this case, but we observe that our best mode is not always the one closest to the published value. 

More specifically, we note that in many of the cases, the mode closest to the published value is not necessarily the most likely or the one whose fit produced the best $\chi^2_\nu$. However, using the full subset of published periods, we find a correlation between the likelihood of the reported modes and the distance to the published value: in $75\%$ of the cases, the closest match is found with the most likely mode, and in $98\%$ of the cases, with one of the three most likely modes; in other words, we establish a $98\%$ confidence that the best solution can be found among the first three most likely modes.

\begin{figure}
    \centering
    \includegraphics[width=\linewidth]{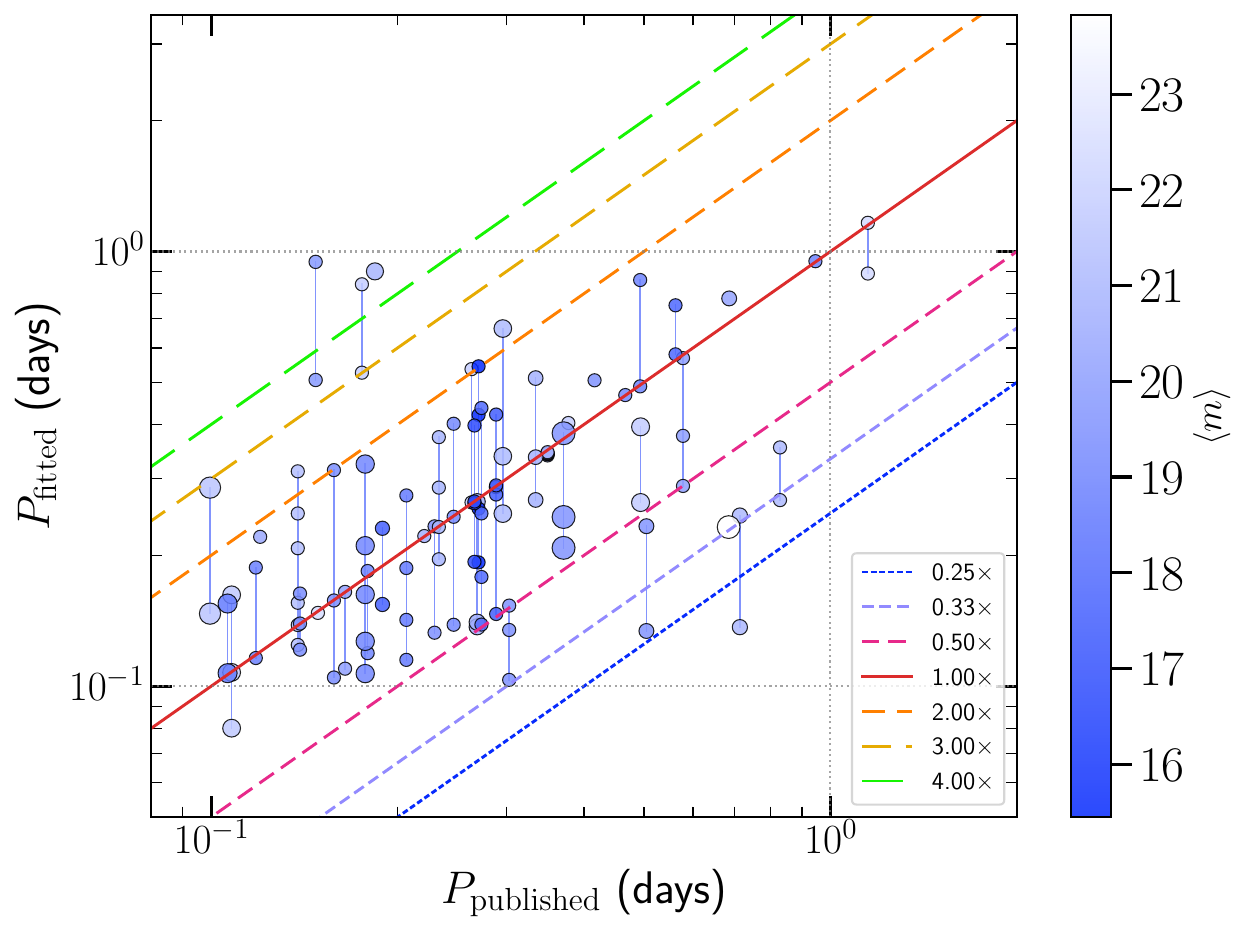}
    \caption{Fitted $P_{\rm ast}$, noted as $P_{\rm fitted}$, against the published $P_{\rm ast}$, noted as $P_{\rm published}$, used for validating the pipeline. In cases presenting period aliases, the different modes are linked by a straight vertical line. The central, continuous red line indicates perfect agreement, while the dashed lines indicate agreement with a harmonic of the published period. The size of the points is proportional to their $\Theta$, where a smaller size implies a better fit. The blue shading of the points indicates their average apparent magnitude, $\langle m \rangle$.}
    \label{fig:fittedreferenceall}
\end{figure}

\begin{figure}
    \centering
    \includegraphics[width=\linewidth]{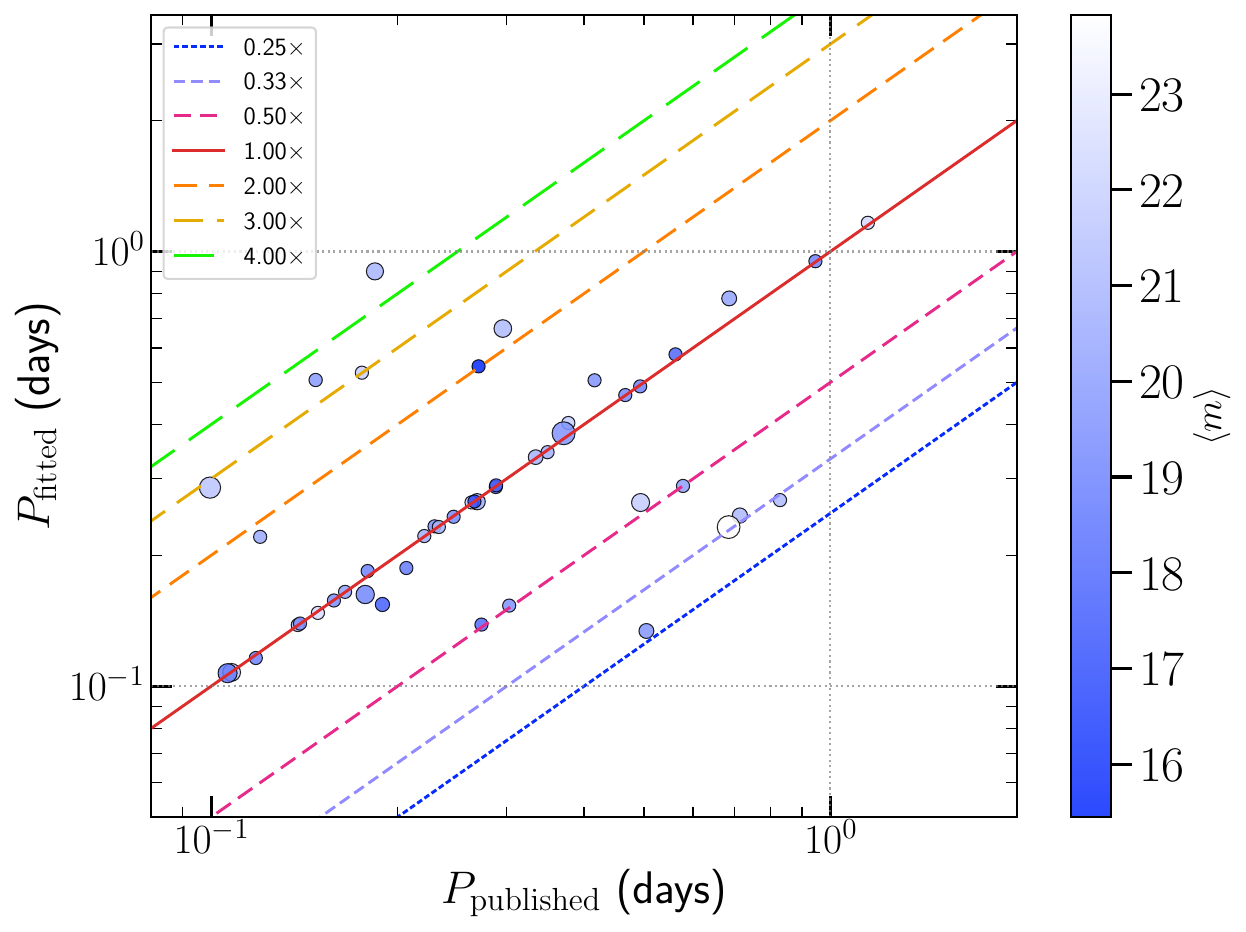}
    \caption{Same as Fig.~\ref{fig:fittedreferenceall} but with the cases presenting aliases collapsed to the alias closest to the published period or its harmonics. }
    \label{fig:fittedreferencecollapsed}
\end{figure}

Figure \ref{fig:fittedreferenceall} shows the location of the fitted periods with respect to their published value. All modes found for a given period are connected by a line. The size of the points is proportional to their $\Theta$, where a smaller size implies a better fit, while a darker shade of blue correlates with a brighter apparent magnitude. Figure \ref{fig:fittedreferencecollapsed} is a version of Fig. \ref{fig:fittedreferenceall} with the modes collapsed to the mode closest to the published value or one of its harmonics. 


\begin{figure}
    \centering
    \includegraphics[width=\linewidth]{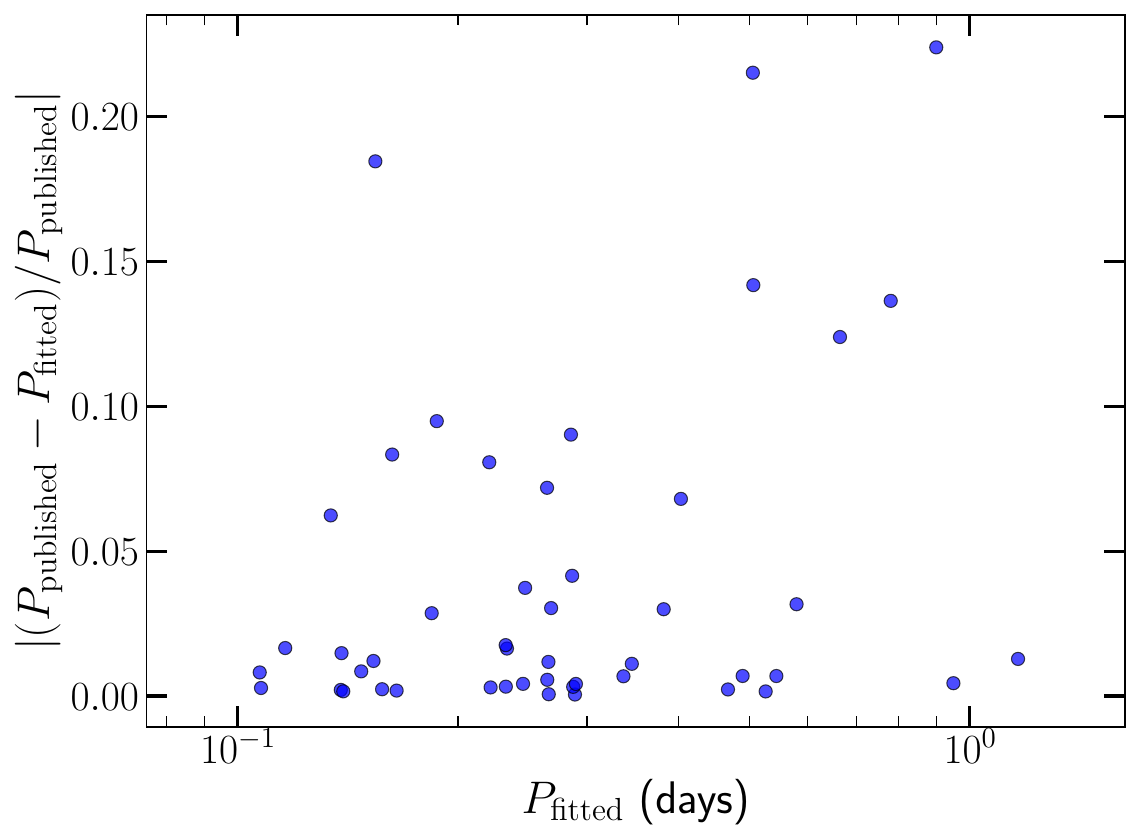}
    \caption{Absolute fractional difference between $P_{\mathrm{published}}$ (or, when applicable, the harmonic of the published period that was used for the comparison) and $P_{\mathrm{fitted}}$ as a function of $P_{\mathrm{fitted}}$.}
    \label{fig:periodvsfractionaldiff}
\end{figure}

After analysing the parameter distribution of the subsample, we find no clear correlation between the length of the fitted period and its divergency from the published value, as is shown in Fig.~\ref{fig:periodvsfractionaldiff}. We note that 1999~JA and 2001~RB$_{71}$, the only two cases with discrepancies greater than $20\%$, also have $\phi_{\rm c}$ values below 0.5, which indicates that they are highly suspect.

\begin{figure}
    \centering
    \includegraphics[width=\linewidth]{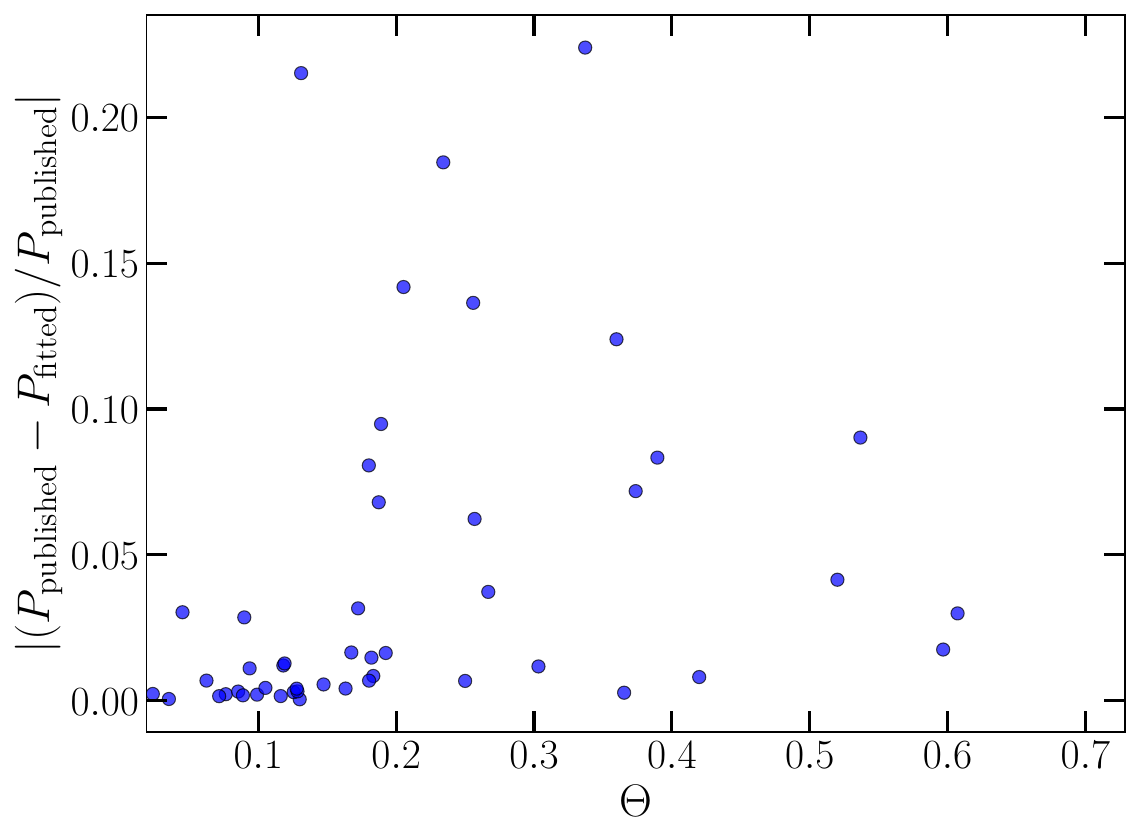}
    \caption{Absolute fractional difference between $P_{\mathrm{published}}$ and $P_{\mathrm{fitted}}$ as a function of $\Theta$.}
    \label{fig:Thetapdm}
\end{figure}

We also inspect the relationship with $\Theta$, shown in Fig.~\ref{fig:Thetapdm}. Although the range of $\Theta$ values in this subset is not wide enough to extract strong conclusions, the plot shows a broad half-trumpet shape, indicating a possible correlation. The outlier to this trend in the top left corner of the figure is one of the objects with low $\phi_{\rm c}$, 1999~JA. The rest of the parameters, such as the number of data points, the relative amplitude of the data, the absolute and apparent magnitude of the asteroid, and the value of $\chi^2_{\nu}$ do not show consistent trends and we did not use them to predict the similarity with the published period.

Taking the published periods as the ground truth, we derived statistics on our accuracy. Looking at the central values, $44\%$ of our periods are within $1\%$ of the published period (or one of its harmonics), $65\%$ within $3\%$, $73\%$ within $5\%$, and $94\%$ within $15\%$. To compare our results with similar previous literature, \cite{lam2023determination} tested their pipeline by using a dataset of 752 published periods and found that $88\%$ of their periods were within $5\%$ of the published ones. However, our dataset reaches much fainter magnitudes and our approach is sensitive to subtle variations (down to 0.1~mag), while their pipeline requires a magnitude variation of at least 0.3~mag to be detectable. The tests of \cite{waszczak2015asteroid} on 927 periods showed that $67\%$ of their periods were within $3\%$ of the published, a comparable result to ours but for brighter magnitudes, centred around 19 mag. Although our validation sample size is smaller than that reported in the literature, we appear to achieve similar accuracy results for much fainter objects. 

In addition, we should note that, even though we set the published periods as the ground truth, in some cases our data from the EES are more densely sampled and have better resolution than the published data using ground-based telescopes. Given that the \Euclid telescope is not affected by atmospheric conditions, it is likely that some of our fitted periods are closer to the true value than those published. Whenever available, the quality code $U$ illustrates the level of confidence of the value published in the literature: $3$ and $3-$ do not allow for other solutions; although unlikely, $2+$ might be in error; $2$ may be wrong by $30\%$, and $2-$ indicates that the coverage was not enough to derive an accurate period. 

Lastly, we rerun our fitting algorithm by adding a second term to the light curve model in Eq. \eqref{eq:single-sine} to assess whether the sinusoidal assumption held. We find that with the additional term, the MCMC routine struggles to converge to reasonable solutions and often overfits the data. This test confirms that in the particular case of our EES data, the simpler single-term model is sufficient. 

\subsection{Period determination for known objects with unknown periods}
\label{subsection:unknown}
\begin{figure}
    \centering
    \includegraphics[width=\linewidth]{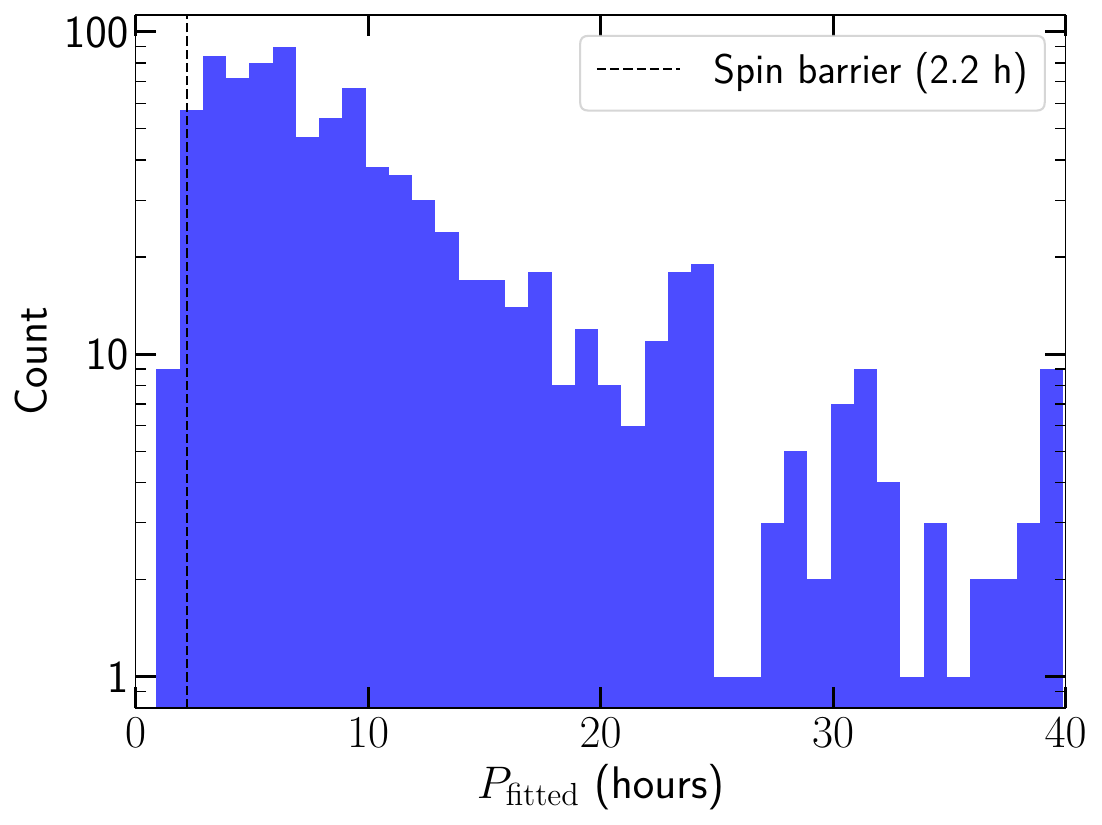}
    \caption{Distribution of all 889 fitted periods in bins of 1\,h. The distribution is denser for periods shorter than 1~day and shows an artificial peak in values near 2~days, resulting from the upper bound of the period search. In cases of multimodality, we plot the most likely period. The dashed black line shows the spin barrier at 0.092~days.}
    \label{fig:periodhistogram}
\end{figure}

\begin{figure}
    \centering
    \includegraphics[width=\linewidth]{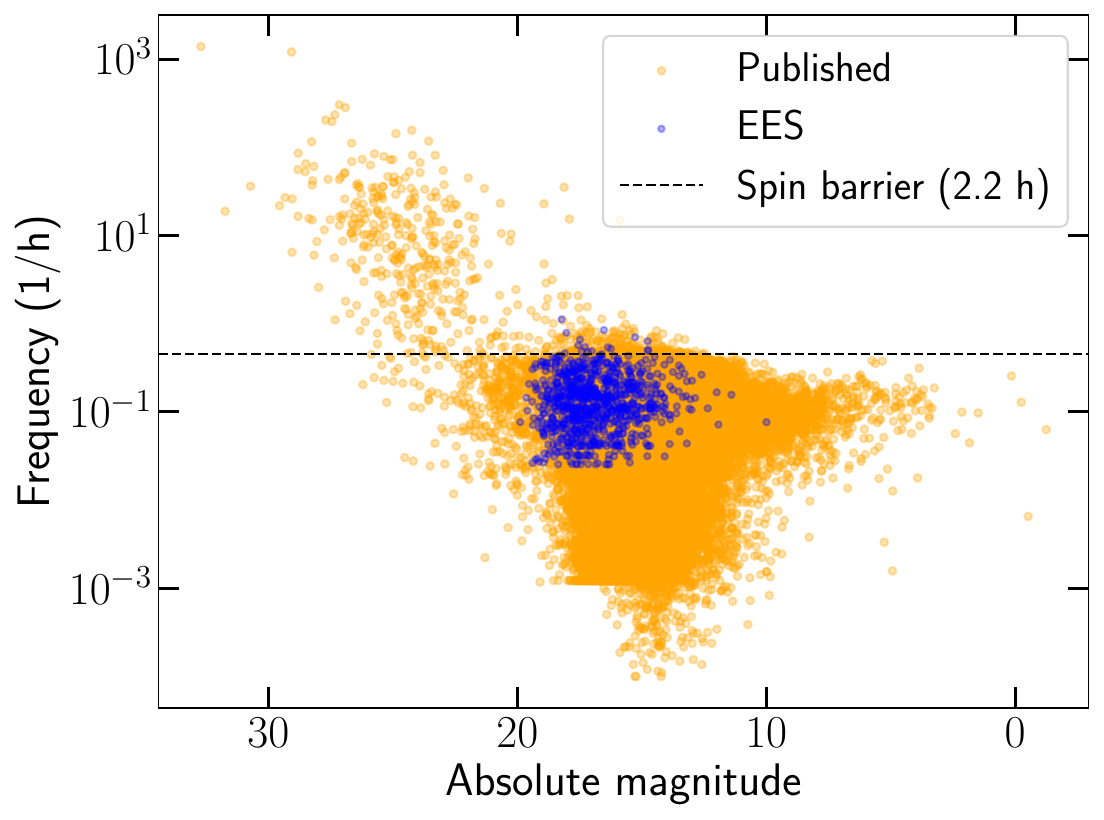}
    \caption{Distribution of all 889 fitted periods in the context of the known distribution, from the literature. The dashed black line shows the spin barrier at 0.092~days.}
    \label{fig:frequencyabsmag}
\end{figure}

\begin{figure}
    \centering
    \includegraphics[width=\linewidth]{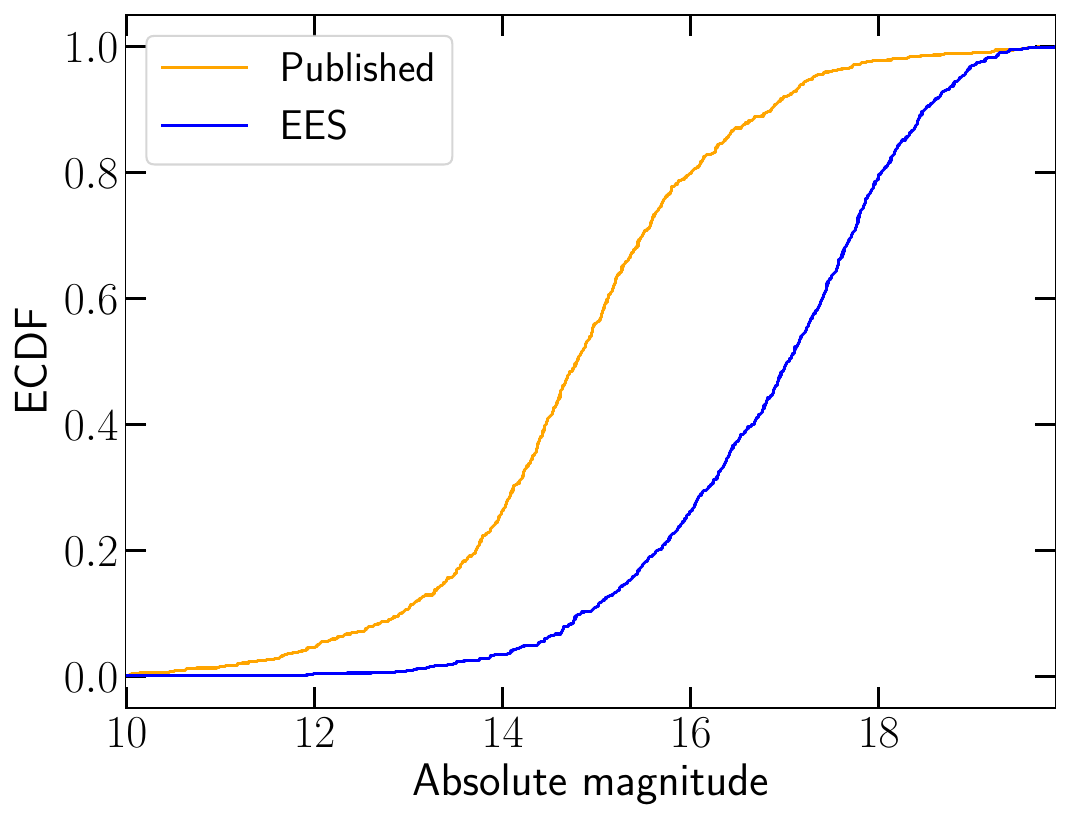}
    \caption{One-dimensional K--S test comparing the empirical cumulative distribution functions (ECDFs) of our magnitude distribution and the published over the same period-magnitude range. The two distributions diverge, with cumulative differences of up to 55\%. As expected, the EES objects are fainter.}
    \label{fig:ecdfmag}
\end{figure}

\begin{figure}
    \centering
    \includegraphics[width=\linewidth]{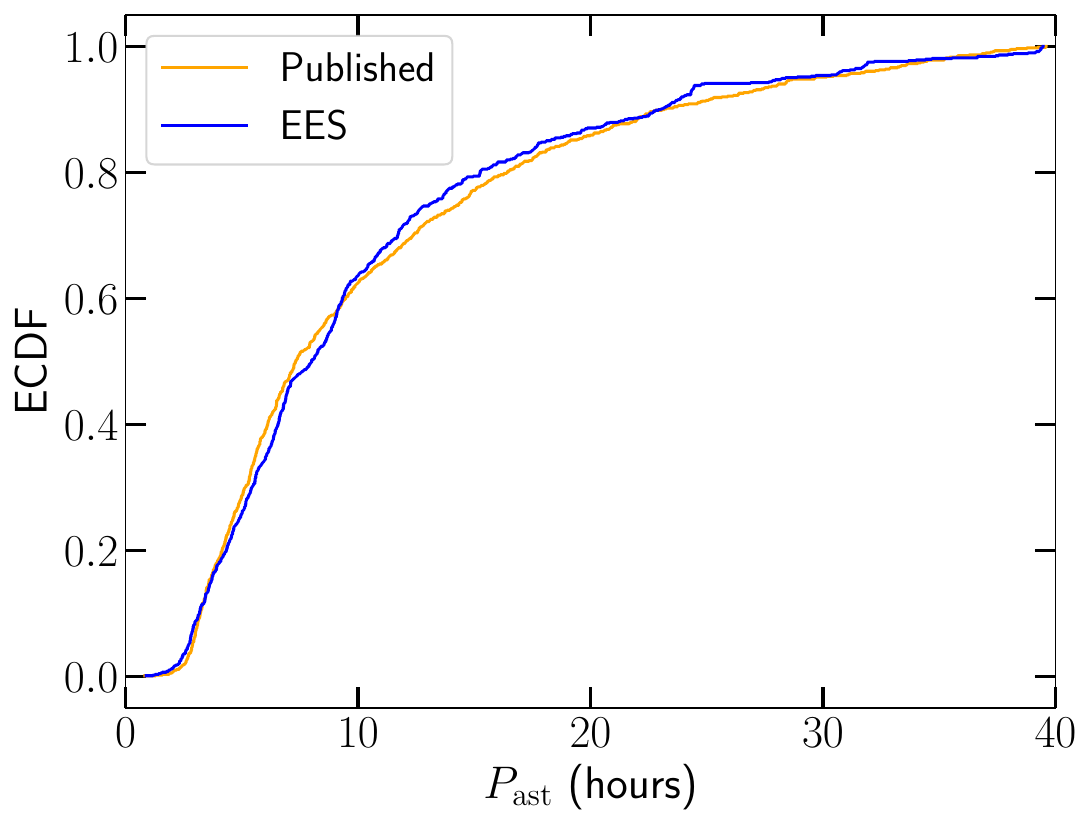}
    \caption{One-dimensional K--S test comparing the empirical cumulative distribution functions (ECDFs) of our period distribution and the published one. The two distributions are found to be broadly consistent, with a maximum difference between the two cumulative distributions of 6\%.}
    \label{fig:ecdf}
\end{figure}

After validating the performance of our pipeline, we use it to analyse the full dataset of 2321 known objects identified in the EES. Following the criteria described in Sect.~\ref{sc:Methods}, we discard 1125 of them because they are unsuited for our period analysis, and 202 after the period finding because they have $\Theta$ above 0.8. Given their size, we expect only a minority of our target asteroids to have periods below the spin barrier of 2.2\,h, which implies that periods found in this range are more likely to be mistaken. We observe that in short periods for which the value of $\Theta$ is high, the algorithm has often fitted the noise. For this reason, we apply an additional filter and discard periods below the spin barrier that have $\Theta$ above 0.5 (see Appendix~\ref{sc:appendix_discarded}). This further removes 105 asteroids from the sample, resulting in a dataset of 889 objects: in this subset, five are Mars crossers, four Cybele, four Hilda, three Hungaria, and 877 are located in other regions of the MB. We present a preview of the results in Table \ref{tab:tableunknown}; the remaining results are available as an electronic table at the Strasbourg Astronomical Data Center (CDS). 

We observe that most of our periods (76\%) present aliases, likely due to insufficient phase coverage caused by the cadence of our exposures. Even though the sampling pattern of the EES is dense, it is highly regular, which may lead to repeated gaps that can result in aliasing. We analyse the phase coverage in our sample and find that the average $\phi_{\rm c}$ is 0.64, with a standard deviation of 0.19. Only a small fraction (9\%) of the periods have full phase coverage, while 15\% of the periods have $\phi_{\rm c}$ below 0.5, which makes them highly suspect. To inform the reader, we include $\phi_{\rm c}$ in Table~\ref{tab:tableunknown}.

The filtered period distribution in hours is shown in Fig. \ref{fig:periodhistogram}, where we can observe that most periods are concentrated in the range below 1~day. The distribution presents a peak in its tail in longer values, artificially caused by the imposed upper bound on the period search, although the underlying periods could be longer. As such, periods near that value are likely longer in reality. We compared our period distribution with the published database using the Kolmogorov--Smirnov (K--S) test \citep{massey1951kolmogorov} and found that the two distributions are compatible, differing by at most 6\% in cumulative fraction, as can be seen in Fig.~\ref{fig:ecdf}. These subtle differences are expected and likely due to selection effects.  

 \begin{figure}
    \centering
    \includegraphics[width=\linewidth]{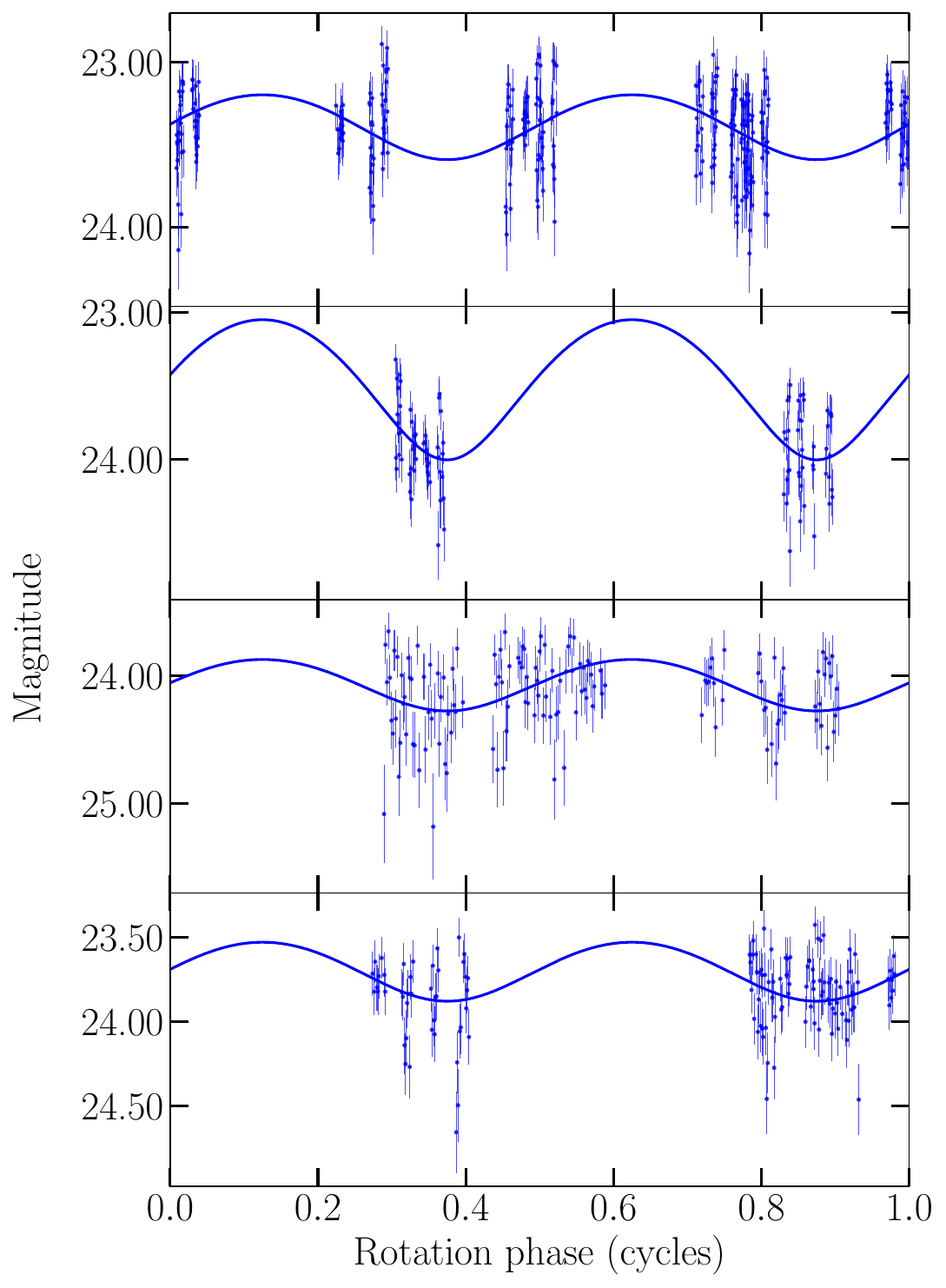}
    \caption{Examples of phase-folded light curves using periods with $\Theta$ above 0.8, indicating a poor solution. All cases with $\Theta$ above 0.8 are discarded from the results. From top to bottom: 1996~RC$_{17}$, 2000~KW$_{84}$, 2001~MR$_{3}$, and 2002~GP$_{193}$}.
    \label{fig:altos}
\end{figure}

\begin{figure}
    \centering
    \includegraphics[width=\linewidth]{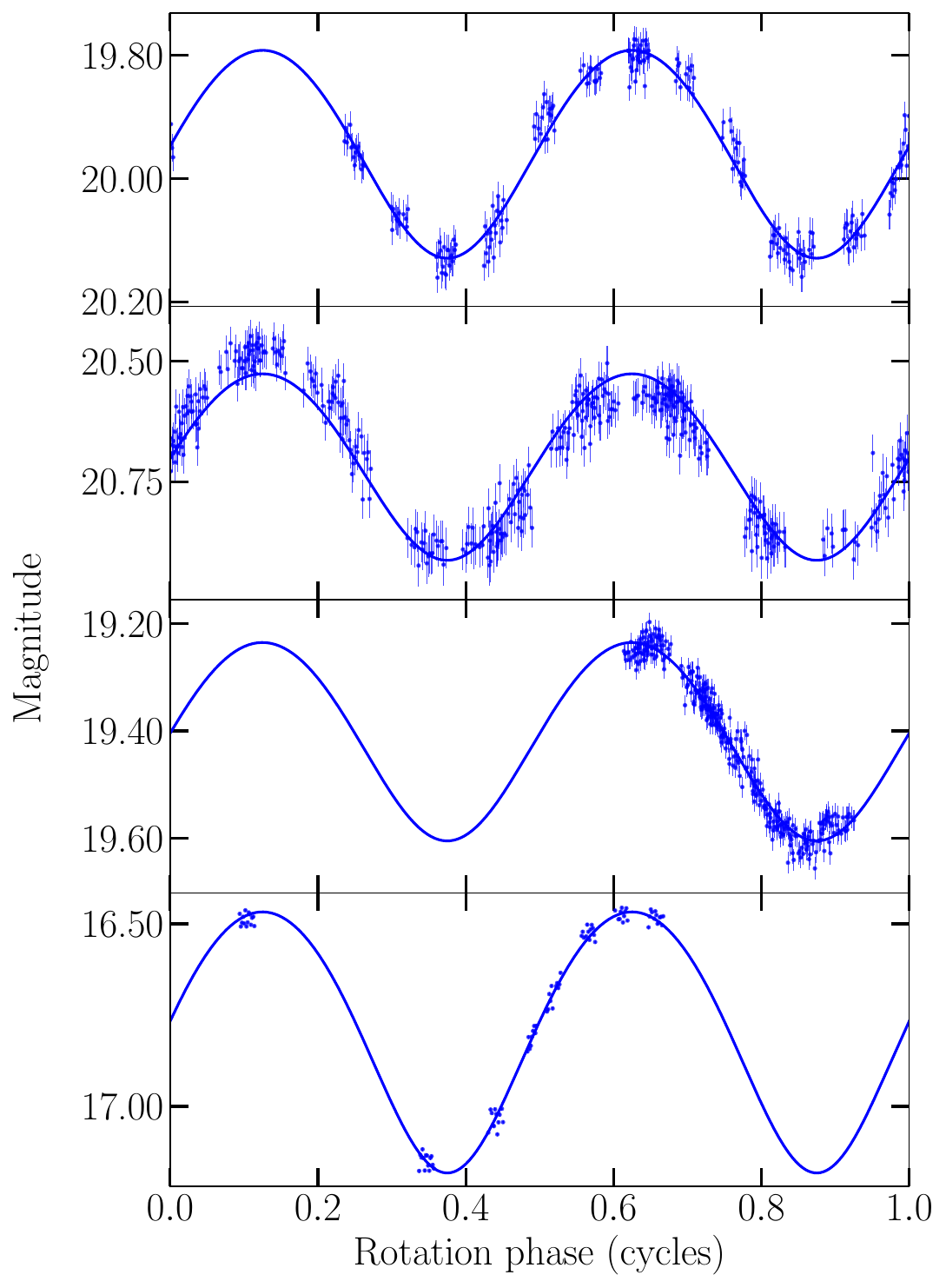}
    \caption{Examples of phase-folded light curves using periods with $\Theta$ below 0.1, indicating an excellent solution. From top to bottom: 2000~VW$_{31}$, 2006~WG$_{108}$, (18957) Mijacobsen, and (2165) Young}.
    \label{fig:bajos}
\end{figure}

\begin{figure}
    \centering
    \includegraphics[width=\linewidth]{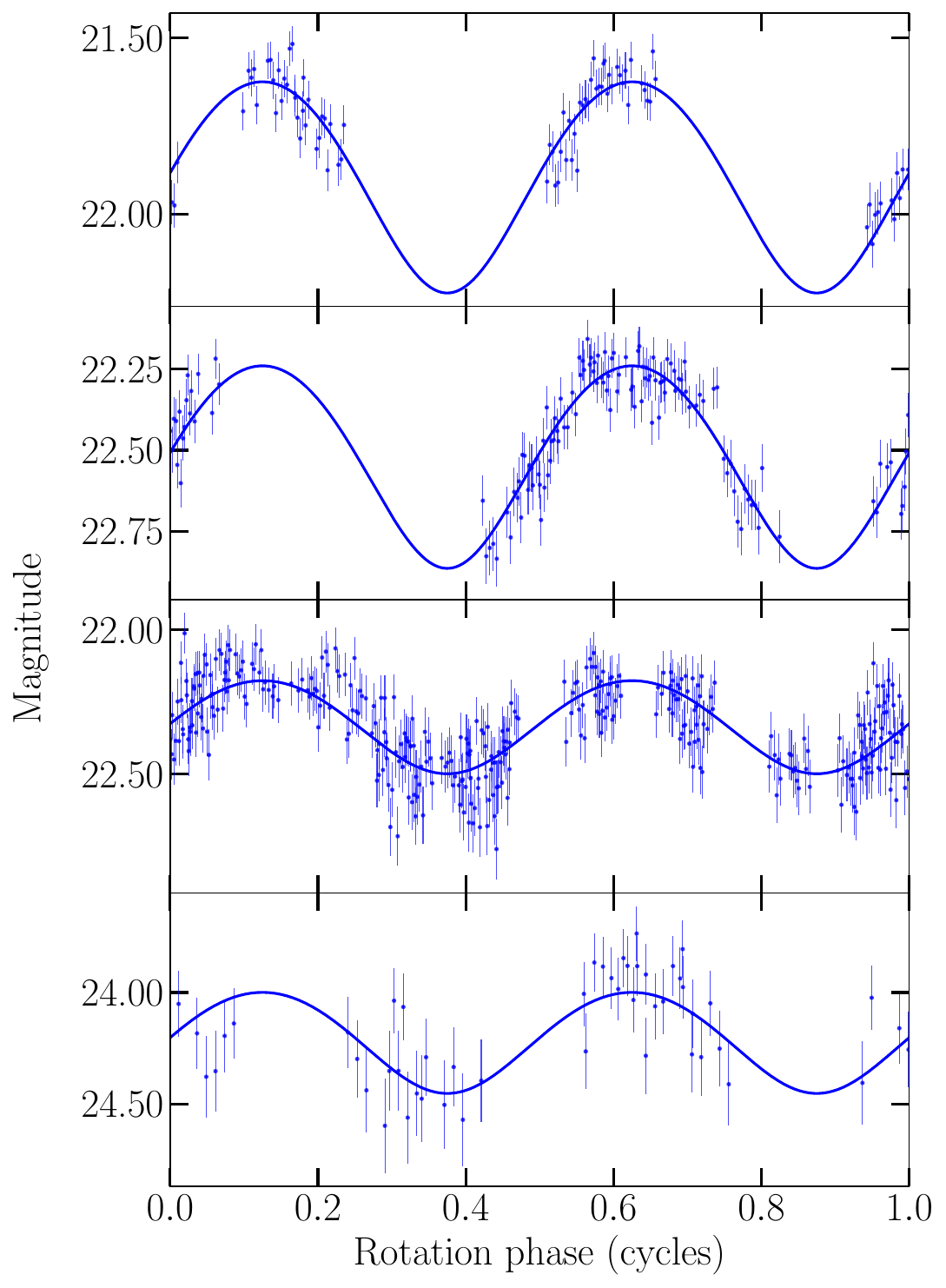}
    \caption{Candidate super-fast rotators, with periods below 2.2\,h. From top to bottom: 2005~SS$_{44}$, 2013~SG$_{64}$, 2015~XS$_{290}$, and 2023~TC$_{192}$}.
    \label{fig:fastrotators}
\end{figure}

We then inspected the distribution of the fitted frequency in our sample as a function of the reported absolute magnitude. We plot the full catalogue of spins in the database in Fig.~\ref{fig:frequencyabsmag}. Similarly to the case of Fig.~\ref{fig:periodhistogram}, we observe an overly dense region in the band corresponding to the upper bound. Most of our asteroids have an absolute magnitude 15--19, which corresponds to a diameter size of approximately 1--5\,km, assuming typical MB distances and albedo. We performed another K--S test to compare our frequency-magnitude distribution to that of the catalogue, and find that the two distributions differ, even after applying the cuts. As can be seen in Fig.~\ref{fig:ecdfmag}, although the distribution of the EES is contained within the bounds of the published distribution, their shapes differ. On average, the objects of the EES are significantly fainter than those published in the catalogue, with cumulative differences of up to 55\%. Given that the \Euclid mission reaches fainter magnitudes than most dedicated asteroid surveys, this behaviour is expected.

The second quality test we performed was to assess the reliability of the $\Theta$ indicator. We visually confirm that whenever $\Theta$ is above 0.8, the fit is almost random and we cannot reliably constrain the period of the asteroid. Figure~\ref{fig:altos} displays several cases of periods with high $\Theta$. The first example shows some structure following the model; however, half of the data points are not aligned with it. The second and fourth examples show signs of variability, but the solutions are clearly not constrained by the data. In the third example, the data are highly dispersed and do not follow the model. We also establish that there is a correlation between low $\Theta$ values and higher-quality periods. In contrast to Fig.~\ref{fig:altos}, Fig.~\ref{fig:bajos} shows four examples with low $\Theta$, all visually accurate. Two thirds ($64\%$) of the fitted periods have $\Theta$ below 0.5.

Within the filtered periods, we find 16 that are below the spin barrier of 2.2\,h. We observe that the light curves in this sample deviate from the first-order sinusoidal shape. This behaviour is expected from the increased rotation speed, which can lead to the presence of binary companions or rotation about an axis other than the principal axis (tumbling). We note that these objects are candidates for being fast rotators since they are below the spin barrier and, given their large size (all below absolute magnitude 19), they are considered candidate super-fast rotators. The rate found of approximately 0.5\% (16/2321) is consistent with the literature \citep{strauss2024decam}. The full list of candidates can be found in Table~\ref{tab:tablesuperfast}, and four example phase-folded light curves are depicted in Fig.~\ref{fig:fastrotators}. The case of 1992~RC$_{4}$, 1999~XN$_{26}$ and 2000~GP$_{60}$ is particularly rare for their large size: 3.0\,km (absolute magnitude 14.8). Since the \Euclid sampling for these asteroids is sparse for such short periods, we will require follow-up observations to further constrain their rotation periods and confirm their status as super-fast rotators.

We built an open database containing the light curves of all 2321 objects studied, all period results, including the 889 high-quality periods, and their most relevant physical parameters. This information will allow all users of our catalogue to interpret and use our data appropriately. 

\section{Conclusions and future work}
\label{sc:Conclusion}

We present a catalogue of spin periods for 889 MB asteroids observed during the EES, which took place at the beginning of the \Euclid mission. This is the first set of spin periods extracted from \Euclid light curves. The photometric precision and long exposures of the VIS images have allowed us to build dense light curves for many objects for which only sparse photometry was available before. Out of the 889 spin periods, $93\%$ (829) had never been reported before. We validate the performance of our pipeline by comparing our fitted periods with the values published in the SsODNet database for the 48 cases with sufficient \Euclid coverage. We find that in $44\%$ of the cases, our fitted period is within $1\%$ of the value published in the literature or one of its harmonics, and that $98\%$ of our periods are within $15\%$ of the published value. For the cases analysed, the largest driver of divergence from the published period is the phase coverage, and other parameters, such as goodness of fit, magnitude, or amplitude, have only a marginal effect. 

With this work, we have aimed to expand our knowledge of the spin-period distribution in the MB and to complement, and even improve, the efforts of ground-based telescopes. Future work will focus on extracting the rotation periods of unknown objects observed during the EES by first linking their streaks across multiple observations. To strengthen our findings, we will incorporate data taken with other surveys. We will also work on upcoming detections from the Euclid Wide Survey, which will provide sparser data but a higher likelihood of encountering near-Earth asteroids. This new dataset will enable us to test our pipeline on a different asteroid population and potentially shorter rotation periods. As the Euclid Wide Survey unfolds during the nominal six-year mission, we will expand our catalogue of asteroid rotational properties in future data releases.

\section*{Data availability}
The data used for this study will become available in November 2026 with the DR1 \citep{DR1cite}. Tables~\ref{tab:tableunknown}, \ref{tab:constant_mag}, and \ref{tab:median_mag} in full are only available in electronic form at the CDS via anonymous ftp to \url{cdsarc.u-strasbg.fr} (130.79.128.5) or via \url{http://cdsweb.u-strasbg.fr/cgi-bin/qcat?J/A+A/}.

\begin{acknowledgements}
We thank Marco Micheli, Odysseas Xenos, and the anonymous reviewer for their helpful comments. This project has received funding from the European Union’s Horizon 2020 and Horizon Europe research and innovation programmes under the Marie Skłodowska-Curie grant agreement Nos 945363 and 101105725, and funding from the Swiss National Science Foundation and the Swiss Innovation Agency (Innosuisse) via the BRIDGE Discovery grant 40B2-0 194729.
  \AckDatalabs
  \AckEC 
\end{acknowledgements}

\bibliography{bibliography}

@misc{DR1cite,
author = "{Euclid Data Release 1}",
howpublished = "\url{https://doi.org/10.57780/esa-d079c51}",
year = 2026
}

@ARTICLE{2018PASP..130f4505T,
       author = {{Tonry}, J.~L. and {Denneau}, L. and {Heinze}, A.~N. and {Stalder}, B. and {Smith}, K.~W. and {Smartt}, S.~J. and {Stubbs}, C.~W. and {Weiland}, H.~J. and {Rest}, A.},
      journal = {PASP},
         year = 2018,
        month = jun,
       volume = {130},
        pages = {064505},
          doi = {10.1088/1538-3873/aabadf},
archivePrefix = {arXiv},
       eprint = {1802.00879},
 primaryClass = {astro-ph.IM}
}

@article{barnouin2019shape,
  author={Barnouin, O.~S. and Daly, M.~G. and Palmer, E.~E. and Gaskell, R.~W. and Weirich, J.~R. and Johnson, C.~L. and Al Asad, M.~M. and Roberts, J.~H. and Perry, M.~E. and Susorney, H.~C.~M. and others},
  journal={Nat. Geosci.},
  volume={12},
  pages={247},
  year={2019}
}

@article{bellm2018zwicky,
  author={Bellm, E.~C. and Kulkarni, S.~R. and Graham, M.~J. and Dekany, R. and Smith, R.~M. and Riddle, R. and Masci, F.~J. and Helou, G. and Prince, T.~A. and Adams, S.~M. and others},
  journal={PASP},
  volume={131},
  pages={018002},
  year={2018}
}

@inproceedings{berthier2006skybot,
  author={Berthier, J. and Vachier, F. and Thuillot, W. and Fernique, P. and Ochsenbein, F. and Genova, F. and Lainey, V. and Arlot, J.-E.},
  booktitle={Astron. Data Anal. Softw. Syst. XV},
  volume={351},
  pages={367},
  year={2006}
}

@article{berthier2023ssodnet,
  author={Berthier, J. and Carry, B. and Mahlke, M. and Normand, J.},
  journal={A\&A},
  volume={671},
  pages={A151},
  year={2023}
}

@article{byrd1995limited,
  author={Byrd, R.~H. and Lu, P. and Nocedal, J. and Zhu, C.},
  journal={SIAM J. Sci. Comput.},
  volume={16},
  pages={1190},
  year={1995}
}

@article{carry2018solar,
  author={Carry, B.},
  journal={A\&A},
  volume={609},
  pages={A113},
  year={2018}
}

@article{cellino2024asteroid,
  author={Cellino, A. and Tanga, P. and Muinonen, K. and Mignard, F.},
  journal={A\&A},
  volume={687},
  pages={A277},
  year={2024}
}

@article{chang2015asteroid,
  author={Chang, C.-K. and Ip, W.-H. and Lin, H.-W. and Cheng, Y.-C. and Ngeow, C.-C. and Yang, T.-C. and Waszczak, A. and Kulkarni, S.~R. and Levitan, D. and Sesar, B. and others},
  journal={ApJS},
  volume={219},
  pages={27},
  year={2015}
}

@article{chang2016large,
  author={Chang, C.-K. and Lin, H.-W. and Ip, W.-H. and Prince, T.~A. and Kulkarni, S.~R. and Levitan, D. and Laher, R. and Surace, J.},
  journal={ApJS},
  volume={227},
  pages={20},
  year={2016}
}

@article{chang2017asteroid,
  author={Chang, C.-K. and Lin, H.-W. and Ip, W.-H. and Prince, T.~A. and Kulkarni, S.~R. and Levitan, D. and Laher, R. and Surace, J.},
  journal={Geoscience Letters},
  volume={4},
  pages={17},
  year={2017}
}

@article{chang2019searching,
  author={Chang, C.-K. and Lin, H.-W. and Ip, W.-H. and Chen, W.-P. and Yeh, T.-S. and Chambers, K.~C. and Magnier, E.~A. and Huber, M.~E. and Flewelling, H.~A. and Waters, C.~Z. and others},
  journal={ApJS},
  volume={241},
  pages={6},
  year={2019}
}

@article{clement2020record,
  author={Clement, M.~S. and Morbidelli, A. and Raymond, S.~N. and Kaib, N.~A.},
  journal={MNRAS},
  volume={492},
  pages={L56},
  year={2020}
}

@inproceedings{conversi2024neomir,
  author={Conversi, L. and Licandro, J. and Delbo, M. and Fitzsimmons, A. and Muinonen, K. and M{\"u}ller, T. and Popescu, M. and Tanga, P. and Moissl, R.},
  booktitle={Space Telescopes and Instrumentation 2024: Optical, Infrared, and Millimeter Wave},
  volume={13092},
  pages={943},
  year={2024},
  organization={SPIE}
}

@article{devogele2024aperture,
  author={Devog{\`e}le, M. and Buzzi, L. and Micheli, M. and Cano, J.~L. and Conversi, L. and Jehin, E. and Ferrais, M. and Oca{\~n}a, F. and F{\"o}hring, D. and Drury, C. and others},
  journal={A\&A},
  volume={689},
  pages={A211},
  year={2024}
}

@article{eglitis2025rotation,
  author={Eglitis, I. and Svincicka, D.},
  journal={Icarus},
  volume={428},
  pages={116380},
  year={2025}
}

@article{erasmus2020investigating,
  author={Erasmus, N. and Navarro-Meza, S. and McNeill, A. and Trilling, D.~E. and Sickafoose, A.~A. and Denneau, L. and Flewelling, H. and Heinze, A. and Tonry, J.~L.},
  journal={ApJS},
  volume={247},
  pages={13},
  year={2020}
}

@article{erasmus2021discovery,
  author={Erasmus, N. and Kramer, D. and McNeill, A. and Trilling, D.~E. and Janse van Rensburg, P. and Van Belle, G.~T. and Tonry, J.~L. and Denneau, L. and Heinze, A. and Weiland, H.~J.},
  journal={MNRAS},
  volume={506},
  pages={3872},
  year={2021}
}

@article{foreman2013emcee,
  author={Foreman-Mackey, D. and Hogg, D.~W. and Lang, D. and Goodman, J.},
  journal={PASP},
  volume={125},
  pages={306},
  year={2013}
}

@article{fraser2016trippy,
  author={Fraser, W. and Alexandersen, M. and Schwamb, M.~E. and Marsset, M. and Pike, R.~E. and Kavelaars, J.~J. and Bannister, M.~T. and Benecchi, S. and Delsanti, A.},
  journal={AJ},
  volume={151},
  pages={158},
  year={2016}
}

@article{garrison2024nifty,
  author={Garrison, L.~H. and Foreman-Mackey, D. and Shih, Y.-H. and Barnett, A.},
  journal={Res. Notes AAS},
  volume={8},
  pages={250},
  year={2024}
}

@inproceedings{gomez2018euclid,
  author={G{\'o}mez-Alvarez, P. and Dupac, X. and Buenadicha, G. and Vavrek, R. and Hoar, J. and Laureijs, R. and Scaramella, R. and Cuillandre, J.~C. and Dinis, J. and Amiaux, J. and others},
  booktitle={Software and Cyberinfrastructure for Astronomy V},
  volume={10707},
  pages={285},
  year={2018},
  organization={SPIE}
}

@article{granvik2017escape,
  author={Granvik, M. and Morbidelli, A. and Vokrouhlick{\`y}, D. and Bottke, W.~F. and Nesvorn{\`y}, D. and Jedicke, R.},
  journal={A\&A},
  volume={598},
  pages={A52},
  year={2017}
}

@article{Greenstreet2026,
  author={Greenstreet, S. and Li, Z. and Vavilov, D.~E. and Singh, D. and Juri\'{c}, M. and Ivezi\'{c}, {\v{Z}}. and Eggl, S. and Koumjian, A. and Moeyens, J. and Carruba, V. and others},
  journal={ApJ},
  volume={996},
  pages={L33},
  year={2026}
}

@article{harris1989photoelectric,
  author={Harris, A.~W. and Young, J.~W. and Bowell, E. and Martin, L.~J. and Millis, R.~L. and Poutanen, M. and Scaltriti, F. and Zappala, V. and Schober, H.~J. and Debehogne, H. and others},
  journal={Icarus},
  volume={77},
  pages={171},
  year={1989}
}

@article{hirabayashi2015internal,
  author={Hirabayashi, M. and S{\'a}nchez, D.~P. and Scheeres, D.~J.},
  journal={ApJ},
  volume={808},
  pages={63},
  year={2015}
}

@inproceedings{kaiser2002pan,
  author={Kaiser, N. and Aussel, H. and Burke, B.~E. and Boesgaard, H. and Chambers, K. and Chun, M.~R. and Heasley, J.~N. and Hodapp, K.-W. and Hunt, B. and Jedicke, R. and others},
  booktitle={Survey and Other Telescope Technologies and Discoveries},
  volume={4836},
  pages={154},
  year={2002},
  organization={SPIE}
}

@article{Kelly2007,
  author={Kelly, B.~C.},
  journal={ApJ},
  volume={665},
  pages={1489},
  year={2007}
}

@article{lagerros1996thermal,
  author={Lagerros, J.~S.~V.},
  journal={A\&A},
  volume={310},
  pages={1011},
  year={1996}
}

@article{lam2023determination,
  author={Lam, A.~L.~H. and Margot, J.-L. and Whittaker, E. and Myhrvold, N.},
  journal={Planet. Sci. J.},
  volume={4},
  pages={61},
  year={2023}
}

@ARTICLE{magnusson1991albedo,
       author = {{Magnusson}, P.},
      journal = {A\&A},
         year = 1991,
        month = mar,
       volume = {243},
        pages = {512}
}

@article{mainzer2023near,
  author={Mainzer, A.~K. and Masiero, J.~R. and Abell, P.~A. and Bauer, J.~M. and Bottke, W. and Buratti, B.~J. and Carey, S.~J. and Cotto-Figueroa, D. and Cutri, R.~M. and Dahlen, D. and others},
  journal={Planet. Sci. J.},
  volume={4},
  pages={224},
  year={2023}
}

@article{marciniak2019thermal,
  author={Marciniak, A. and Ali-Lagoa, V. and M{\"u}ller, T.~G. and Szak{\'a}ts, R. and Moln{\'a}r, L. and P{\'a}l, A. and Podlewska-Gaca, E. and Parley, N. and Antonini, P. and Barbotin, E. and others},
  journal={A\&A},
  volume={625},
  pages={A139},
  year={2019}
}

@article{marzari2020evolution,
  author={Marzari, F. and Rossi, A. and Golubov, O. and Scheeres, D.~J.},
  journal={AJ},
  volume={160},
  pages={128},
  year={2020}
}

@article{massey1951kolmogorov,
  author={Massey, F.~J., Jr.},
  journal={J. Am. Stat. Assoc.},
  volume={46},
  pages={68},
  year={1951}
}

@misc{mccully2018astroscrappy,
  author={McCully, C. and Crawford, S. and Kovacs, G. and Tollerud, E. and Betts, E. and Bradley, L. and Craig, M. and Turner, J. and Streicher, O. and Sipocz, B. and Robitaille, T. and Deil, C.},
  title={astropy/astroscrappy: v1.0.5 Zenodo Release},
  year={2018},
  publisher={Zenodo},
  doi={10.5281/zenodo.1482019},
  url={https://doi.org/10.5281/zenodo.1482019}
}

@article{mcneill2023untargeted,
  author={McNeill, A. and Gowanlock, M. and Mommert, M. and Trilling, D.~E. and Llama, J. and Paddock, N.},
  journal={AJ},
  volume={166},
  pages={152},
  year={2023}
}

@article{muller1998asteroids,
  author={M{\"u}ller, T.~G. and Lagerros, J.~S.~V.},
  journal={A\&A},
  volume={338},
  pages={340},
  year={1998}
}

@article{nucita2025euclid,
  author={Nucita, A.~A. and Conversi, L. and Verdier, A. and Franco, A. and Sacquegna, S. and P{\"o}ntinen, M. and Altieri, B. and Carry, B. and De Paolis, F. and Strafella, F. and others},
  journal={A\&A},
  volume={694},
  pages={A116},
  year={2025}
}

@article{ostro1995radar,
  author={Ostro, S.~J. and Hudson, R.~S. and Jurgens, R.~F. and Rosema, K.~D. and Campbell, D.~B. and Yeomans, D.~K. and Chandler, J.~F. and Giorgini, J.~D. and Winkler, R. and Rose, R. and others},
  journal={Sci},
  volume={270},
  pages={80},
  year={1995}
}

@article{pal2020solar,
  author={P{\'a}l, A. and Szak{\'a}ts, R. and Kiss, C. and B{\'o}di, A. and Bogn{\'a}r, Z. and Kalup, C. and Kiss, L.~L. and Marton, G. and Moln{\'a}r, L. and Plachy, E. and others},
  journal={ApJS},
  volume={247},
  pages={26},
  year={2020}
}

@article{persson2022stability,
  author={Persson, B.~N.~J. and Biele, J.},
  journal={Tribol. Lett.},
  volume={70},
  pages={34},
  year={2022}
}

@article{podlewska2021determination,
  author={Podlewska-Gaca, E. and Poleski, R. and Bartczak, P. and McDonald, I. and P{\'a}l, A.},
  journal={ApJS},
  volume={255},
  pages={4},
  year={2021}
}

@article{polishook2012asteroid,
  author={Polishook, D. and Ofek, E.~O. and Waszczak, A. and Kulkarni, S.~R. and Gal-Yam, A. and Aharonson, O. and Laher, R. and Surace, J. and Klein, C. and Bloom, J. and others},
  journal={MNRAS},
  volume={421},
  pages={2094},
  year={2012}
}

@article{pontinen2020euclid,
  author={P{\"o}ntinen, M. and Granvik, M. and Nucita, A.~A. and Conversi, L. and Altieri, B. and Auricchio, N. and Bodendorf, C. and Bonino, D. and Brescia, M. and Capobianco, V. and others},
  journal={A\&A},
  volume={644},
  pages={A35},
  year={2020}
}

@techreport{pontinen2025asteroid,
  author={P{\"o}ntinen, M. and Granvik, M. and Nucita, A. and Sacquegna, S. and Franco, A. and Carry, B. and Chang, B.~Y.~I.-G.},
  title={Asteroid Science with ESA Euclid: Results from the Ecliptic PDC Campaign},
  year={2025},
  institution={Copernicus Meetings}
}

@article{pravec2002asteroid,
  author={Pravec, P. and Harris, A.~W. and Michalowski, T.},
  journal={Asteroids III},
  volume={113},
  pages={35},
  year={2002}
}

@article{pravec2006photometric,
  author={Pravec, P. and Scheirich, P. and Ku{\v{s}}nir{\'a}k, P. and {\v{S}}arounov{\'a}, L. and Mottola, S. and Hahn, G. and Brown, P. and Esquerdo, G. and Kaiser, N. and Krzeminski, Z. and others},
  journal={Icarus},
  volume={181},
  pages={63},
  year={2006}
}

@article{rozitis2011directional,
  author={Rozitis, B. and Green, S.~F.},
  journal={MNRAS},
  volume={415},
  pages={2042},
  year={2011}
}

@article{Rubincam2000,
  author={Rubincam, D.~P.},
  journal={Icarus},
  volume={148},
  pages={2},
  year={2000}
}

@article{scheeres2006dynamical,
  author={Scheeres, D.~J. and Fahnestock, E.~G. and Ostro, S.~J. and Margot, J.-L. and Benner, L.~A.~M. and Broschart, S.~B. and Bellerose, J. and Giorgini, J.~D. and Nolan, M.~C. and Magri, C. and others},
  journal={Sci},
  volume={314},
  pages={1280},
  year={2006}
}

@article{sergeyev2025rotation,
  author={Sergeyev, A.~V. and Carry, B. and Eggl, S. and Berthier, J. and Santerne, A. and Vachier, F. and Shevchenko, V.~G.},
  journal={A\&A},
  volume={703},
  pages={A302},
  year={2025}
}

@article{Spencer1989,
  author={Spencer, J.~R. and Lebofsky, L.~A. and Sykes, M.~V.},
  journal={Icarus},
  volume={78},
  pages={337},
  year={1989}
}

@article{stellingwerf1978period,
  author={Stellingwerf, R.~F.},
  journal={ApJ},
  volume={224},
  pages={953},
  year={1978}
}

@article{Stephens2012b,
  author={Stephens, R.~D.},
  journal={Minor Planet Bull.},
  volume={39},
  pages={80},
  year={2012}
}

@article{strauss2024decam,
  title={The DECam Ecliptic Exploration Project (DEEP). VII. The Strengths of Three Superfast Rotating Main-belt Asteroids from a Preliminary Search of DEEP Data},
  author={Strauss, Ryder and McNeill, Andrew and Trilling, David E and Valdes, Francisco and Bernardinelli, Pedro H and Fuentes, Cesar and Gerdes, David W and Holman, Matthew J and Juri{\'c}, Mario and Lin, Hsing Wen and others},
  journal={AJ},
  volume={168},
  number={4},
  pages={184},
  year={2024},
  publisher={The American Astronomical Society}
}

@article{szabo2022rotation,
  author={Szab{\'o}, G.~M. and P{\'a}l, A. and Szigeti, L. and Bogn{\'a}r, Z. and B{\'o}di, A. and Kalup, C. and J{\"a}ger, Z.~J. and Kiss, L.~L. and Kiss, C. and Kov{\'a}cs, J. and others},
  journal={A\&A},
  volume={661},
  pages={A48},
  year={2022}
}

@article{taylor2007spin,
  author={Taylor, P.~A. and Margot, J.-L. and Vokrouhlicky, D. and Scheeres, D.~J. and Pravec, P. and Lowry, S.~C. and Fitzsimmons, A. and Nolan, M.~C. and Ostro, S.~J. and Benner, L.~A.~M. and others},
  journal={Sci},
  volume={316},
  pages={274},
  year={2007}
}

@article{vanderplas2018understanding,
  author={VanderPlas, J.~T.},
  journal={ApJS},
  volume={236},
  pages={16},
  year={2018}
}

@article{vavilov2025rotation,
  author={Vavilov, D.~E. and Carry, B.},
  journal={A\&A},
  volume={693},
  pages={A66},
  year={2025}
}

@article{vdurech2018asteroid,
  author={{\v{D}}urech, J. and Hanu{\v{s}}, J. and Ali-Lagoa, V.},
  journal={A\&A},
  volume={617},
  pages={A57},
  year={2018}
}

@article{vdurech2022rotation,
  author={{\v{D}}urech, J. and V{\'a}vra, M. and Van{\v{c}}o, R. and Erasmus, N.},
  journal={Front. Astron. Space Sci.},
  volume={9},
  pages={809771},
  year={2022}
}

@article{vdurech2023reconstruction,
  author={{\v{D}}urech, J. and Hanu{\v{s}}, J.},
  journal={A\&A},
  volume={675},
  pages={A24},
  year={2023}
}

@article{virtanen2016streak,
  author={Virtanen, J. and Poikonen, J. and S{\"a}ntti, T. and Komulainen, T. and Torppa, J. and Granvik, M. and Muinonen, K. and Pentik{\"a}inen, H. and Martikainen, J. and N{\"a}r{\"a}nen, J. and others},
  journal={Adv. Space Res.},
  volume={57},
  pages={1607},
  year={2016}
}

@article{waszczak2015asteroid,
  author={Waszczak, A. and Chang, C.-K. and Ofek, E.~O. and Laher, R. and Masci, F. and Levitan, D. and Surace, J. and Cheng, Y.-C. and Ip, W.-H. and Kinoshita, D. and others},
  journal={AJ},
  volume={150},
  pages={75},
  year={2015}
}

@INPROCEEDINGS{wiktorowicz2017albedo,
       author = {{Wiktorowicz}, S. and {Masiero}, J.~R.},
    booktitle = {AAS/Division for Planetary Sciences Meeting Abstracts},
       series = {AAS/Division for Planetary Sciences Meeting Abstracts},
       volume = {49},
        pages = {110.32},
         year = 2017
}

@article{wright2010wide,
  author={Wright, E.~L. and Eisenhardt, P.~R.~M. and Mainzer, A.~K. and Ressler, M.~E. and Cutri, R.~M. and Jarrett, T. and Kirkpatrick, J.~D. and Padgett, D. and McMillan, R.~S. and Skrutskie, M. and others},
  journal={AJ},
  volume={140},
  pages={1868},
  year={2010}
}

@article{yeh2020asteroid,
  author={Yeh, T.-S. and Li, B. and Chang, C.-K. and Zhao, H.-B. and Ji, J.-H. and Lin, Z.-Y. and Ip, W.-H.},
  journal={AJ},
  volume={160},
  pages={73},
  year={2020}
}

@ARTICLE{Q1-TP002,
       author = {{Euclid Collaboration: McCracken}, H.~J. and {Benson}, K. and {Dolding}, C. and others},
        title = "{Euclid Quick Data Release (Q1): VIS processing and data products}",
      journal = {A\&A, in press (Euclid Q1 SI), \url{https://doi.org/10.1051/0004-6361/202554594}},
         year = 2026,
        month = mar,
          eid = {arXiv:2503.15303},
        pages = {arXiv:2503.15303},
archivePrefix = {arXiv},
       eprint = {2503.15303},
 primaryClass = {astro-ph.IM},
       adsurl = {https://ui.adsabs.harvard.edu/abs/2025arXiv250315303E}
}

@ARTICLE{EuclidSkyOverview,
author = {{Euclid Collaboration: Mellier}, Y. and {Abdurro'uf} and {Acevedo~Barroso}, J.A. and others},
	title = {Euclid - I. Overview of the Euclid mission},
	DOI= "10.1051/0004-6361/202450810",
	url= "https://doi.org/10.1051/0004-6361/202450810",
	journal = {A\&A},
	year = 2025,
	volume = 697,
	pages = "A1",
}

@ARTICLE{EuclidSkyVIS,
author = {{Euclid Collaboration: Cropper}, M. and {Al-Bahlawan}, A. and {Amiaux}, J. and others},
	title = {Euclid - II. The VIS instrument},
	DOI= "10.1051/0004-6361/202450996",
	url= "https://doi.org/10.1051/0004-6361/202450996",
	journal = {A\&A},
	year = 2025,
	volume = 697,
	pages = "A2",
}

@ARTICLE{EuclidSkyNISP,
author = {{Euclid Collaboration: Jahnke}, K. and {Gillard}, W. and {Schirmer}, M. and others},
	title = {Euclid - III. The NISP Instrument},
	DOI= "10.1051/0004-6361/202450786",
	url= "https://doi.org/10.1051/0004-6361/202450786",
	journal = {A\&A},
	year = 2025,
	volume = 697,
	pages = "A3",
}

@ARTICLE{Scaramella-EP1,
       author = {{Euclid Collaboration: Scaramella}, R. and {Amiaux}, J. and {Mellier}, Y. and others},
        title = "{Euclid preparation. I. The Euclid Wide Survey}",
      journal = {\aap},
         year = 2022,
        month = jun,
       volume = {662},
          eid = {A112},
        pages = {A112},
          doi = {10.1051/0004-6361/202141938},
archivePrefix = {arXiv},
       eprint = {2108.01201},
 primaryClass = {astro-ph.CO},
       adsurl = {https://ui.adsabs.harvard.edu/abs/2022A&A...662A.112E}
}

\begin{appendix}
\onecolumn
\section{Tables}
We present in Table~\ref{tab:tableknown} a comparison between our fitted periods and those reported in the SSODNet catalogue; in Table~\ref{tab:tableunknown}, a preview of the fitted periods from the EES, and in Table~\ref{tab:tablesuperfast}, the candidate super-fast rotators in the EES. For a detailed discussion, see Sect.~\ref{subsection:Validation} and Sect.~\ref{subsection:unknown}.

\begin{longtable}{
  p{0.12\textwidth} @{}
  p{0.12\textwidth}
  p{0.12\textwidth}
  p{0.12\textwidth}
  p{0.07\textwidth}
  p{0.06\textwidth}
  p{0.08\textwidth}
  p{0.10\textwidth}
  p{0.05\textwidth}
}
\caption{Comparison between our fitted periods and those reported in the SSODNet catalogue. }\label{tab:tableknown} \\
\toprule
Name &
$P_{\mathrm{fitted}}$ [h] &
$P_{\mathrm{published}}$ or harmonic [h] &
$P_{\mathrm{published}}$ [h] &
Harmonic factor &
Quality code &
Fractional difference &
$\langle m \rangle$ [mag] &
$\phi_{\rm c}$ \\
\midrule
\endfirsthead

\multicolumn{9}{c}%
{\tablename\ \thetable\ -- Continued from previous page} \\
\toprule
Name &
$P_{\mathrm{fitted}}$ [h] &
$P_{\mathrm{published}}$ or harmonic [h] &
$P_{\mathrm{published}}$ [h] &
Harmonic factor &
Quality code &
Fractional difference &
$\langle m \rangle$ [mag] &
$\phi_{\rm c}$ \\
\midrule
\endhead

\midrule
\multicolumn{9}{r}{Continued on next page} \\
\endfoot

\bottomrule
\endlastfoot
1993 FD$_{22}$ & $6.93^{+0.02}_{-0.02}$ & $6.94^{+0.01}_{-0.01}$ & $13.88^{+0.02}_{-0.02}$ & 1/2 & U=2 & $0.001$ & $20.04^{+0.01}_{-0.01}$ & $0.3$ \\[1em]
1995 CD & $4.49^{+0.01}_{-0.01}$ & $4.960^{+0.003}_{-0.003}$ & $4.960^{+0.003}_{-0.003}$ & 1 & U=2 & $0.095$ & $18.67^{+0.01}_{-0.01}$ & $0.4$ \\[1em]
1997 BQ$_{6}$ & $6.35^{+0.03}_{-0.02}$ & $5.92^{+0.14}_{-0.14}$ & $11.85^{+0.29}_{-0.29}$ & 1/2 & U=2 & $0.072$ & $21.824^{+0.008}_{-0.008}$ & $0.8$ \\[1em]
1998 BB$_{34}$ & $3.679^{+0.008}_{-0.007}$ & $3.64$ & $7.27$ & 1/2 &  & $0.012$ & $19.530^{+0.002}_{-0.002}$ & $0.5$ \\[1em]
1998 FB$_{61}$ & $5.889^{+0.004}_{-0.004}$ & $5.9136^{+0.0001}_{-0.0001}$ & $5.9136^{+0.0001}_{-0.0001}$ & 1 & U=2 & $0.004$ & $19.192^{+0.002}_{-0.002}$ & $0.8$ \\[1em]
1998 SU$_{134}$ & $8.29^{+0.04}_{-0.03}$ & $8.39$ & $8.39$ & 1 & U=2 & $0.011$ & $20.69^{+0.03}_{-0.03}$ & $0.4$ \\[1em]
1999 JA & $12.134^{+0.008}_{-0.003}$ & $9.99^{+0.08}_{-0.08}$ & $9.99^{+0.08}_{-0.08}$ & 1 &  & $0.215$ & $19.54^{+0.01}_{-0.01}$ & $0.3$ \\[1em]
2000 AY$_{239}$ & $5.30^{+0.03}_{-0.02}$ & $5.76^{+0.06}_{-0.06}$ & $2.88^{+0.03}_{-0.03}$ & 2 & U=2$-$ & $0.081$ & $20.488^{+0.003}_{-0.003}$ & $1.0$ \\[1em]
2000 CZ$_{45}$ & $11.22^{+0.04}_{-0.04}$ & $11.1948^{+0.0001}_{-0.0001}$ & $11.1948^{+0.0001}_{-0.0001}$ & 1 &  & $0.002$ & $19.191^{+0.002}_{-0.002}$ & $0.6$ \\[1em]
2000 GX$_{75}$ & $5.93^{+0.16}_{-0.11}$ & $5.71$ & $17.14$ & 1/3 &  & $0.037$ & $21.075^{+0.007}_{-0.005}$ & $0.9$ \\[1em]
2000 KY$_{8}$ & $6.87^{+0.09}_{-0.09}$ & $7.17^{+0.06}_{-0.06}$ & $2.39^{+0.02}_{-0.02}$ & 3 & U=2 & $0.040$ & $21.51^{+0.03}_{-0.03}$ & $0.6$ \\[1em]
2000 QG$_{120}$ & $5.60^{+0.01}_{-0.01}$ & $5.51$ & $5.51$ & 1 &  & $0.016$ & $19.478^{+0.002}_{-0.002}$ & $0.8$ \\[1em]
2001 FQ$_{77}$ & $6.38^{+0.07}_{-0.07}$ & $6.4548^{+0.0001}_{-0.0001}$ & $6.4548^{+0.0001}_{-0.0001}$ & 1 &  & $0.012$ & $20.57^{+0.02}_{-0.02}$ & $0.4$ \\[1em]
2001 OF$_{8}$ & $3.216^{+0.005}_{-0.005}$ & $3.03$ & $12.11$ & 1/4 &  & $0.062$ & $19.697^{+0.002}_{-0.002}$ & $1.0$ \\[1em]
2001 QB$_{93}$ & $18.72^{+0.40}_{-0.37}$ & $16.47$ & $16.47$ & 1 &  & $0.136$ & $20.255^{+0.005}_{-0.005}$ & $0.3$ \\[1em]
2001 RB$_{71}$ & $21.61^{+0.12}_{-0.21}$ & $17.65348^{+0.00004}_{-0.00004}$ & $4.41337^{+0.00001}_{-0.00001}$ & 4 &  & $0.224$ & $20.92^{+0.03}_{-0.03}$ & $0.4$ \\[1em]
2001 VJ$_{6}$ & $15.96^{+0.05}_{-3.40}$ & $14.20^{+0.40}_{-0.40}$ & $7.10^{+0.20}_{-0.20}$ & 2 & U=2+ & $0.124$ & $21.14^{+0.01}_{-0.01}$ & $0.6$ \\[1em]
2002 JJ$_{74}$ & $6.85^{+0.07}_{-0.03}$ & $6.28^{+0.24}_{-0.24}$ & $1.57^{+0.06}_{-0.06}$ & 4 & U=2$-$ & $0.090$ & $21.16^{+0.02}_{-0.02}$ & $0.4$ \\[1em]
2002 WP$_{14}$ & $5.32^{+0.03}_{-0.02}$ & $5.30$ & $5.30$ & 1 &  & $0.003$ & $20.676^{+0.006}_{-0.006}$ & $0.6$ \\[1em]
2003 AE$_{31}$ & $5.579^{+0.004}_{-0.004}$ & $5.60^{+0.01}_{-0.01}$ & $5.60^{+0.01}_{-0.01}$ & 1 & U=2 & $0.003$ & $20.905^{+0.005}_{-0.005}$ & $0.4$ \\[1em]
2003 BD$_{76}$ & $3.319^{+0.002}_{-0.002}$ & $3.31$ & $3.31$ & 1 &  & $0.002$ & $21.284^{+0.008}_{-0.007}$ & $0.4$ \\[1em]
2004 FT$_{110}$ & $3.54^{+0.01}_{-0.01}$ & $3.57^{+0.02}_{-0.02}$ & $3.57^{+0.02}_{-0.02}$ & 1 & U=2 & $0.009$ & $22.452^{+0.006}_{-0.006}$ & $0.6$ \\[1em]
2004 LF$_{4}$ & $6.43^{+0.01}_{-0.02}$ & $6.63$ & $19.90$ & 1/3 &  & $0.030$ & $21.079^{+0.005}_{-0.007}$ & $0.8$ \\[1em]
2005 GV$_{127}$ & $12.63^{+0.05}_{-0.06}$ & $12.61$ & $4.20$ & 3 &  & $0.002$ & $21.869^{+0.005}_{-0.005}$ & $0.8$ \\[1em]
2005 WD$_{10}$ & $6.36^{+0.01}_{-0.01}$ & $6.32^{+0.08}_{-0.08}$ & $6.32^{+0.08}_{-0.08}$ & 1 & U=2 & $0.006$ & $21.852^{+0.007}_{-0.008}$ & $0.7$ \\[1em]
2006 TV$_{74}$ & $9.68^{+0.06}_{-0.06}$ & $9.06^{+0.17}_{-0.17}$ & $9.06^{+0.17}_{-0.17}$ & 1 & U=2 & $0.068$ & $21.852^{+0.007}_{-0.007}$ & $0.8$ \\[1em]
2014 AH$_{7}$ & $27.94^{+0.38}_{-0.37}$ & $27.59^{+0.47}_{-0.47}$ & $27.59^{+0.47}_{-0.47}$ & 1 & U=2 & $0.013$ & $22.38^{+0.01}_{-0.01}$ & $0.6$ \\[1em]
2016 UR$_{100}$ & $5.58^{+0.07}_{-0.07}$ & $5.48^{+0.18}_{-0.18}$ & $16.44^{+0.54}_{-0.54}$ & 1/3 & U=2 & $0.018$ & $23.83^{+0.02}_{-0.02}$ & $0.6$ \\[1em]
Ashkinazi & $6.895^{+0.004}_{-0.003}$ & $6.92$ & $6.92$ & 1 &  & $0.003$ & $18.515^{+0.001}_{-0.002}$ & $0.9$ \\[1em]
Bowenyueli & $8.08^{+0.01}_{-0.01}$ & $8.022^{+0.003}_{-0.003}$ & $8.022^{+0.003}_{-0.003}$ & 1 & U=2 & $0.007$ & $20.76^{+0.01}_{-0.02}$ & $0.5$ \\[1em]
Cornaa & $12.15^{+0.17}_{-0.17}$ & $14.16^{+0.16}_{-0.16}$ & $3.54^{+0.04}_{-0.04}$ & 4 & U=2 & $0.142$ & $19.819^{+0.003}_{-0.003}$ & $0.6$ \\[1em]
Corradolam. & $22.80^{+0.14}_{-0.13}$ & $22.702^{+0.001}_{-0.001}$ & $22.702^{+0.001}_{-0.001}$ & 1 &  & $0.004$ & $18.959^{+0.002}_{-0.002}$ & $0.5$ \\[1em]
Gismonda & $13.07^{+11.40}_{-0.05}$ & $12.98^{+0.01}_{-0.01}$ & $6.488^{+0.005}_{-0.005}$ & 2 & U=3 & $0.007$ & $15.447^{+0.001}_{-0.002}$ & $0.7$ \\[1em]
Handley & $13.92^{+0.03}_{-0.03}$ & $13.49^{+0.84}_{-0.84}$ & $13.49^{+0.84}_{-0.84}$ & 1 &  & $0.032$ & $17.895^{+0.001}_{-0.001}$ & $0.6$ \\[1em]
Haydn & $2.787^{+0.006}_{-0.006}$ & $2.83$ & $2.83$ & 1 &  & $0.017$ & $18.895^{+0.001}_{-0.002}$ & $0.7$ \\[1em]
Hiraimasa & $6.95^{+0.01}_{-0.01}$ & $6.93$ & $6.93$ & 1 &  & $0.004$ & $17.508^{+0.003}_{-0.003}$ & $0.6$ \\[1em]
Jayaranjan & $11.750^{+0.009}_{-0.009}$ & $11.83$ & $11.83$ & 1 &  & $0.007$ & $18.964^{+0.002}_{-0.002}$ & $0.6$ \\[1em]
McCauley & $3.701^{+0.008}_{-0.008}$ & $4.54$ & $4.54$ & 1 &  & $0.185$ & $17.588^{+0.001}_{-0.001}$ & $0.7$ \\[1em]
Medkeff & $2.58^{+0.01}_{-0.01}$ & $2.59^{+0.01}_{-0.01}$ & $2.59^{+0.01}_{-0.01}$ & 1 & U=2 & $0.003$ & $21.697^{+0.005}_{-0.005}$ & $0.8$ \\[1em]
Mijacobsen & $3.780^{+0.002}_{-0.002}$ & $3.7890^{+0.0007}_{-0.0007}$ & $3.7890^{+0.0007}_{-0.0007}$ & 1 & U=2 & $0.002$ & $19.404^{+0.002}_{-0.002}$ & $0.4$ \\[1em]
Noctua & $4.42^{+4.08}_{-0.03}$ & $4.2953^{+0.0002}_{-0.0002}$ & $4.2953^{+0.0002}_{-0.0002}$ & 1 &  & $0.029$ & $19.00^{+0.14}_{-0.00}$ & $0.4$ \\[1em]
Okabayashi & $2.57^{+0.01}_{-0.01}$ & $2.5515^{+0.0005}_{-0.0005}$ & $2.5515^{+0.0005}_{-0.0005}$ & 1 & U=3$-$ & $0.008$ & $18.651^{+0.002}_{-0.002}$ & $0.9$ \\[1em]
Oliver & $3.327^{+0.001}_{-0.001}$ & $3.28^{+0.05}_{-0.05}$ & $6.56^{+0.10}_{-0.10}$ & 1/2 &  & $0.015$ & $17.994^{+0.001}_{-0.001}$ & $0.7$ \\[1em]
Pecorelli & $3.345^{+0.002}_{-0.002}$ & $3.34$ & $3.34$ & 1 & U=3$-$ & $0.002$ & $19.369^{+0.003}_{-0.003}$ & $0.6$ \\[1em]
Reynolds & $3.957^{+0.006}_{-0.006}$ & $3.94923^{+0.00001}_{-0.00001}$ & $3.94923^{+0.00001}_{-0.00001}$ & 1 &  & $0.002$ & $20.156^{+0.004}_{-0.004}$ & $0.5$ \\[1em]
Vespucci & $9.17^{+0.21}_{-0.18}$ & $8.90$ & $8.90$ & 1 &  & $0.030$ & $19.616^{+0.007}_{-0.007}$ & $0.6$ \\[1em]
Young & $6.385^{+0.003}_{-0.003}$ & $6.389^{+0.003}_{-0.003}$ & $6.389^{+0.003}_{-0.003}$ & 1 &  & $0.001$ & $16.7670^{+0.0009}_{-0.0009}$ & $0.6$ \\[1em]
Yuuko & $3.90^{+0.01}_{-0.01}$ & $4.26^{+0.14}_{-0.14}$ & $4.26^{+0.14}_{-0.14}$ & 1 &  & $0.083$ & $19.122^{+0.002}_{-0.002}$ & $0.5$ \\[1em]
\end{longtable}
\tablefoot{Due to space constraints, we do not include the numbers of named asteroids.}

\vspace{1 cm}

\begin{table}[H]
\centering
\caption{Fitted periods from the EES (preview).}
\label{tab:tableunknown}
\begin{tabular}{
  p{0.17\textwidth} @{}
  p{0.07\textwidth}
  p{0.06\textwidth}
  p{0.09\textwidth}
  p{0.09\textwidth}
  p{0.10\textwidth}
  p{0.06\textwidth}
  p{0.06\textwidth}
  p{0.05\textwidth}
  p{0.06\textwidth}
}
\toprule
Name & $P_1$ [h] & $(A_{\rm p-p})_1$ [mag] & $P_2$ [h] & $P_3$ [h] & $\langle m \rangle$ [mag] & $N_{\rm data points}$ & $\chi^2_{\nu}$ & $\phi_{\rm c}$ & $\Theta$ \\
\midrule
(39418) 1204 T-2 &
$2.885^{+0.008}_{-0.008}$ & 0.17 &
$4.72^{+0.02}_{-0.03}$ &
$2.083^{+0.003}_{-0.004}$ &
$20.565^{+0.003}_{-0.004}$ &
154 & 2.78 & 1.0 & 0.34 \\[1.4em]
1981 EL$_{43}$ &
$3.017^{+0.007}_{-0.007}$ & 0.13 &
$4.99^{+0.03}_{-0.02}$ &
 &
$20.863^{+0.002}_{-0.002}$ &
177 & 1.02 & 1.0 & 0.24 \\[1.4em]
1981 EO$_{12}$ &
$2.910^{+0.006}_{-0.006}$ & 0.24 &
$4.67^{+0.02}_{-0.02}$ &
 &
$21.146^{+0.003}_{-0.003}$ &
161 & 1.04 & 1.0 & 0.17 \\[1.4em]
\bottomrule
\end{tabular}
\tablefoot{The first three aliases, $P_1$, $P_2$, and $P_3$, are presented in order of likelihood.}
\end{table}

\onecolumn
\begin{longtable}{
  p{0.14\textwidth} @{}
  p{0.07\textwidth}
  p{0.06\textwidth}
  p{0.08\textwidth}
  p{0.08\textwidth}
  p{0.10\textwidth}
  p{0.06\textwidth}
  p{0.06\textwidth}
  p{0.04\textwidth}
  p{0.04\textwidth}
  p{0.04\textwidth}
}
\caption{Candidate super-fast rotators in the EES.}
\label{tab:tablesuperfast} \\
\toprule
Name & $P_1$ [h] & $(A_{\rm p-p})_1$ [mag] & $P_2$ [h] & $P_3$ [h] & $\langle m \rangle$ [mag] & $N_{\rm data points}$ & $\chi^2_{\nu}$ & $\phi_{\rm c}$ & $\Theta$ & $H$ \\
\midrule
\endfirsthead

\multicolumn{11}{c}%
{\tablename\ \thetable\ -- Continued from previous page} \\
\toprule
Name & $P_1$ [h] & $(A_{\rm p-p})_1$ [mag] & $P_2$ [h] & $P_3$ [h] & $\langle m \rangle$ [mag] & $N_{\rm data points}$ & $\chi^2_{\nu}$ & $\phi_{\rm c}$ & $\Theta$ & $H$ \\
\midrule
\endhead

\midrule
\multicolumn{11}{r}{Continued on next page} \\
\endfoot

\bottomrule
\endlastfoot

1992 RC$_{4}$ &
$1.97^{+0.02}_{-0.01}$ & 0.12 &
$2.239^{+0.009}_{-0.008}$ &
$1.746^{+0.009}_{-0.009}$ &
$19.914^{+0.003}_{-0.003}$ &
50 & 0.86 & 0.5 & 0.30 & 14.75 \\[1.4em]

1999 XN$_{26}$ &
$1.577^{+0.004}_{-0.004}$ & 0.19 &
$1.97^{+1.14}_{-0.01}$ &
$1.300^{+0.003}_{-0.004}$ &
$20.385^{+0.004}_{-0.004}$ &
64 & 1.96 & 1.0 & 0.24 & 14.76 \\[1.4em]

2000 EZ$_{46}$ &
$1.422^{+0.006}_{-0.006}$ & 0.06 &
$1.57^{+0.01}_{-0.01}$ &
$1.113^{+0.004}_{-0.005}$ &
$20.071^{+0.003}_{-0.002}$ &
142 & 1.05 & 1.0 & 0.47 & 15.27 \\[1.4em]

2000 GP$_{60}$ &
$2.04^{+0.00}_{-0.01}$ & 0.13 &
 &
 &
$19.443^{+0.003}_{-0.003}$ &
67 & 1.86 & 0.6 & 0.26 & 14.75 \\[1.4em]

2001 SL$_{197}$ &
$2.093^{+0.009}_{-0.008}$ & 0.30 &
$1.420^{+0.004}_{-0.004}$ &
$1.111^{+0.004}_{-0.005}$ &
$22.43^{+0.01}_{-0.01}$ &
70 & 1.10 & 0.8 & 0.39 & 17.17 \\[1.4em]

2003 VN$_{7}$ &
$1.179^{+0.003}_{-0.002}$ & 0.13 &
$1.023^{+0.002}_{-0.002}$ &
 &
$21.491^{+0.005}_{-0.005}$ &
134 & 1.05 & 1.0 & 0.45 & 16.52 \\[1.4em]

2005 GF$_{169}$ &
$2.06^{+0.69}_{-0.01}$ & 0.25 &
$1.80^{+0.02}_{-0.02}$ &
 &
$22.49^{+0.01}_{-0.01}$ &
89 & 1.01 & 0.8 & 0.34 & 17.41 \\[1.4em]

2005 SS$_{44}$ &
$1.750^{+0.002}_{-0.002}$ & 0.60 &
$1.895^{+0.003}_{-0.003}$ &
$2.065^{+0.002}_{-0.003}$ &
$21.88^{+0.02}_{-0.02}$ &
85 & 1.13 & 0.6 & 0.44 & 17.55 \\[1.4em]

2006 FW$_{39}$ &
$1.510^{+0.008}_{-0.007}$ & 0.19 &
$1.91^{+0.01}_{-0.01}$ &
$1.273^{+0.003}_{-0.003}$ &
$22.287^{+0.007}_{-0.008}$ &
165 & 1.11 & 0.9 & 0.42 & 17.48 \\[1.4em]

2012 OD$_{4}$ &
$2.052^{+0.003}_{-0.003}$ & 0.36 &
$1.88^{+0.03}_{-0.05}$ &
$2.043^{+0.003}_{-0.003}$ &
$21.902^{+0.008}_{-0.007}$ &
80 & 1.09 & 0.6 & 0.27 & 17.01 \\[1.4em]

2012 TV$_{96}$ &
$1.854^{+0.006}_{-0.007}$ & 0.26 &
$3.69^{+0.02}_{-0.02}$ &
$2.44^{+0.02}_{-0.02}$ &
$22.58^{+0.01}_{-0.01}$ &
70 & 1.13 & 0.6 & 0.38 & 16.36 \\[1.4em]

2013 SG$_{64}$ &
$1.83^{+1.62}_{-0.00}$ & 0.63 &
$1.221^{+0.002}_{-0.002}$ &
 &
$22.51^{+0.04}_{-0.01}$ &
133 & 1.03 & 0.7 & 0.22 & 17.28 \\[1.4em]

2014 WR$_{154}$ &
$1.269^{+0.002}_{-0.002}$ & 0.38 &
$1.080^{+0.002}_{-0.002}$ &
 &
$23.24^{+0.01}_{-0.01}$ &
111 & 1.02 & 1.0 & 0.45 & 18.03 \\[1.4em]

2015 XS$_{290}$ &
$2.156^{+0.003}_{-0.003}$ & 0.32 &
$2.995^{+0.005}_{-0.005}$ &
 &
$22.326^{+0.005}_{-0.005}$ &
359 & 1.01 & 1.0 & 0.37 & 18.09 \\[1.4em]

2023 TC$_{192}$ &
$0.893^{+0.003}_{-0.003}$ & 0.45 &
$0.843^{+0.003}_{-0.002}$ &
$0.799^{+0.003}_{-0.003}$ &
$24.20^{+0.02}_{-0.03}$ &
51 & 1.06 & 0.8 & 0.48 & 18.22 \\[1.4em]

(41450) Medkeff &
$1.922^{+0.008}_{-0.008}$ & 0.12 &
$3.90^{+0.03}_{-0.03}$ &
$2.58^{+0.01}_{-0.01}$ &
$21.697^{+0.005}_{-0.005}$ &
111 & 1.04 & 0.8 & 0.37 & 16.24 \\[1.4em]

\end{longtable}
\tablefoot{The first three aliases, $P_1$, $P_2$, and $P_3$, are presented in order of likelihood. The absolute magnitude, $H$, is taken from SsODNet and can be used as a size proxy.}

\section{Discarded light curves}
We present in Table~\ref{tab:constant_mag} a preview of the light curves that did not show enough variability in the observed time range to allow for period determination. A description of the methods used for this filtering can be found in Sect.~\ref{subsection:period_determination}. We present in Table~\ref{tab:median_mag} period fits that we considered low quality and excluded from the sample of 889 high-quality periods, as is discussed in Sect.~\ref{subsection:performance}.
\label{sc:appendix_discarded}
\begin{longtable}{
  p{0.22\textwidth} @{}
  p{0.18\textwidth}
  p{0.14\textwidth}
  p{0.14\textwidth}
  p{0.10\textwidth}
  p{0.10\textwidth}
}
\caption{Light curves from the EES that did not show variability (preview).}
\label{tab:constant_mag} \\
\toprule
Name & $\langle m \rangle$ [mag] & $\sigma_m$ [mag] & $N_{\rm data points}$ & Range [h] & $\chi^2_{\nu}$ \\
\midrule
\endfirsthead

\multicolumn{6}{c}%
{\tablename\ \thetable\ -- Continued from previous page} \\
\toprule
Name & $\langle m \rangle$ [mag] & $\sigma_m$ [mag] & $N_{\rm data points}$ & Range [h] & $\chi^2_{\nu}$ \\
\midrule
\endhead

\midrule
\multicolumn{6}{r}{Continued on next page} \\
\endfoot

\bottomrule
\endlastfoot

1992 BT$_{5}$  & 20.76 & 0.04 & 112 & 8.21 & 1.92 \\[1.2em]
1993 BR$_{11}$  & 23.03 & 0.11 & 166 & 15.69 & 1.97 \\[1.2em]
1993 FZ$_{52}$  & 21.95 & 0.06 & 211 & 19.73 & 1.61 \\[1.2em]
1994 SE$_{7}$  & 21.77 & 0.04 & 63 & 12.14 & 1.47 \\[1.2em]
1994 FJ$_{9}$  & 20.18 & 0.02 & 157 & 15.98 & 1.02 \\[1.2em]  
\end{longtable}
\tablefoot{The value of the fitted constant is displayed as $\langle m \rangle$.}

\vspace{1 cm}

\begin{longtable}{
  p{0.22\textwidth} @{}
  p{0.18\textwidth}
  p{0.14\textwidth}
  p{0.14\textwidth}
  p{0.10\textwidth}
  p{0.10\textwidth}
}
\caption{Periods from the EES that were considered low-quality fits (preview).}
\label{tab:median_mag} \\
\toprule
Name & $\langle m \rangle$ [mag] & $\sigma_m$ [mag] & $N_{\rm data points}$ & Range [h] & $\Theta$ \\
\midrule
\endfirsthead

\multicolumn{6}{c}%
{\tablename\ \thetable\ -- Continued from previous page} \\
\toprule
Name & $\langle m \rangle$ [mag] & $\sigma_m$ [mag] & $N_{\rm data points}$ & Range [h] & $\Theta$ \\
\midrule
\endhead

\midrule
\multicolumn{6}{r}{Continued on next page} \\
\endfoot

\bottomrule
\endlastfoot

1996 RC$_{17}$ & $23.38^{+0.01}_{-0.01}$ & 0.24 & 309 & 19.74 & 0.85 \\[1.2em]
2000 KW$_{84}$ & $23.4^{+0.2}_{-0.2}$ & 0.27 & 92 & 8.51 & 0.85 \\[1.2em]
2001 MR$_{3}$ & $24.05^{+0.02}_{-0.02}$ & 0.30 & 139 & 4.77 & 0.89 \\[1.2em]
2002 GP$_{193}$ & $23.69^{+0.04}_{-0.04}$ & 0.22 & 131 & 15.90 & 0.85 \\[1.2em]
2002 TP$_{330}$ & $23.42^{+0.05}_{-0.07}$ & 0.16 & 89 & 4.78 & 0.92 \\[1.2em]

\end{longtable}

\section{Sources of known periods used}
\label{sc:used_periods}

We present the sources of the periods used for comparison with our results in Sect.~\ref{sc:Results}.\\

\cite{Stephens2012b}; \cite{chang2015asteroid}; \cite{waszczak2015asteroid}; \cite{chang2016large}; \cite{vdurech2018asteroid}; \cite{chang2019searching}; \cite{erasmus2020investigating}; \cite{pal2020solar}; \cite{yeh2020asteroid}; \cite{podlewska2021determination}; \cite{vdurech2022rotation}; \cite{vdurech2023reconstruction}; \cite{lam2023determination}; \cite{mcneill2023untargeted}; \cite{cellino2024asteroid}.

\end{appendix}

\label{LastPage}
\end{document}